\documentclass[journal,10pt,onecolumn,draftclsnofoot,]{IEEEtran}
\usepackage[pdftex]{graphicx}
\usepackage{cite}
\usepackage{amssymb}
\usepackage{amsmath}
\usepackage{graphics}
\usepackage{graphicx}
\usepackage{float}
\usepackage[T1]{fontenc}
\usepackage[utf8]{inputenc}
\usepackage{authblk}
\usepackage{authblk}
\usepackage{tabularx}
\usepackage{balance}
\usepackage{ amssymb }
\usepackage{algorithm2e}
\usepackage{algorithmic}
\usepackage{mathtools}
\usepackage{gensymb}
\usepackage{subfig}
\usepackage{multicol}
\usepackage{lipsum}
\usepackage{chngpage}
\usepackage{calc}
\usepackage{ragged2e}
\usepackage{tabularx}
\usepackage{balance}

\begin{document}

\title{Amplitude, Phase and Quadrant (APQ) Modulation for Indoor  Visible Light Communications}

\author{~Hanaa~Abumarshoud$^{\dagger}$,~Lina~Mohjazi$^{\ast}$~and~Sami~Muhaidat$^{\ddagger}$\\
$^{\dagger}$LiFi~R$\&$D~Centre,~University~of~Strathclyde,~Glasgow,~UK\\ (e-mail:~hanaa.abumarshoud@strath.ac.uk) \\
$^{\ast}$School~of~Engineering,~University~of~Glasgow,~Glasgow,~UK\\ (e-mail:~l.mohjazi@ieee.org) \\
$^{\ddagger}$Center~for~Cyber-Physical~Systems,~Department~of~Electrical~and~Computer~Engineering,\\~Khalifa~University,~Abu~Dhabi,~UAE \\ (e-mail: muhaidat@ieee.org)\\
}
\maketitle

\begin{abstract}
The main challenge in visible light communications (VLC) is the low modulation bandwidth of light-emitting diodes (LEDs). This forms a barrier towards achieving high data rates. Moreover, the implementation of high order modulation schemes is restricted by the requirements of intensity modulation (IM) and direct detection (DD), which demand the use of real unipolar signals. In this paper, we propose a novel  amplitude, phase and quadrant (APQ) modulation scheme that fits into the IM/DD restrictions in VLC systems. The proposed scheme decomposes the complex and bipolar symbols of high order modulations  into three different symbols that carry the amplitude, phase and quadrant information of the intended symbol. The constructed symbols are assigned different power levels and are transmitted simultaneously, i.e. exploiting the entire bandwidth and time resources. The receiving terminal performs successive interference cancellation to extract and decode the three different symbols, and then uses them to decide the intended complex bipolar symbol. We evaluate the performance of the proposed APQ scheme in terms of symbol-error-rate and achievable system throughput for different setup scenarios. The obtained results are compared with generalized spatial shift keying (GSSK). The presented results show that APQ offers a higher reliability compared to GSSK across the simulation area, while providing lower hardware complexity.
\end{abstract}

\section{Introduction}
The ever increasing demand for high-speed wireless data connectivity has motivated  researchers to look beyond the conventional radio frequency (RF) communications. As a result, the wireless communication industry started moving  toward using the radio spectrum above $10$ GHz, i.e.,  mmWave communications, to cope with the influx in data traffic. However, the increase in path loss at such high frequencies necessitates the employment of small cells with strong line-of-sight (LOS) paths.  Nevertheless, it is a challenging task to provide  backhaul infrastructures to support the deployment of mmWave  small cells.

Visible light communication (VLC) has emerged as a promising candidate to support and complement conventional RF communications. Specifically, VLC uses light-emitting diodes (LEDs) as small cells to provide wireless connectivity to a small number of users over a short distance of a few meters. To this end, the intensity of the LED transmitted light is modulated to convey the information signal. This process is known as intensity modulation (IM). At the receiving terminal, a photo detector (PD) is employed to perform direct detection (DD) by  translating the fluctuations in the received light intensity into an electrical current  that is used for data demodulation. The requirements for the modification of existing lighting infrastructure to support VLC are far simpler and cheaper than the deployment of new infrastructure to support new communication standards. Hence, VLC is considered  a potential  compelling technology for supporting conventional RF communications \cite{7360112}. Huge research interests have been directed towards the integration of VLC systems in heterogeneous networks for ubiquitous connectivity. In this regards, VLC can offer exceptionally high data rates \cite{7275086}, highly secure communications \cite{9070153} and seamless multi-user access \cite{8713381}.      However, the realization of the full potentials of VLC is subject to its  ability to provide sufficiently high data rates, this is particularly crucial due to the following factors: 1) the limited modulation bandwidth of the currently used phosphorescent white LEDs, which spans a few MHz, and 2) the constraints of IM/DD that require the transmitted signal to be positive and real, hindering the implementation of high order modulation schemes. Consequently, the deployment of high spectral efficiency modulation techniques that fit into the constraints of  IM/DD is critical in the design of high data-rates VLC systems.

Various high spectral efficiency modulation techniques  have been proposed for VLC.  For instance, multi-carrier modulation  by means of orthogonal frequency-division multiplexing (OFDM) has been widely considered for downlink VLC systems \cite{8955888,9139190}. Since  OFDM signals are inherently bipolar and complex, modifications to the conventional OFDM technique are needed to fit into the constraints of IM/DD. To satisfy the reality constraint, Hermitian symmetry is applied on the parallel data streams into the IFFT input in OFDM modules, leading to a spectral loss of half of the available bandwidth. Moreover, to satisfy the non-negativity constraint, a DC bias is added to the generated multicarrier waveform, leading to higher  peak-to-average power ratio and increased sensitivity  to the LED non-linearity. Several approaches have been proposed for reducing the required DC bias in VLC OFDM systems, including DC-clipped OFDM (DCO-OFDM) \cite{DCO3, DCO4, DCO5, DCO6}, asymmetrically clipped optical OFDM (ACO-OFDM)\cite{ACO1,Aco2},  asymmetrically clipped DC biased optical OFDM (ADO-OFDM) \cite{ADO-OFDM} and Unipolar OFDM (U-OFDM) \cite{eU-OFDM}. Nevertheless, such modifications come at the  cost of additional processing complexity \cite{OFDMlimitation}.

Space shift keying (SSK) has been proposed as a low complexity modulation technique that is less prone to the LED non-linearity compared to OFDM. SSK is a multiple-input technique which  uses the spatial dimension to transmit data \cite{SM1}. In conventional SSK,  only one transmitting LED is activated at any symbol duration, such that the spatial position of the transmitting LED determines the transmitted symbol \cite{SSK1}. The spectral efficiency of SSK has been improved by proposing generalized SSK (GSSK) \cite{GSSK1,GSSK3}. In this scheme, more than one transmitting LEDs are activated  at any symbol duration, such that $2^{N_T}$ possible  combinations of transmitters are used to generate a spatial symbol of $N_T$ bits, where $N_T$ denotes the total number of LEDs. The performance of GSSK, however, is highly dependent on the dissimilarity among the  channel gains of different receivers. This requirement forms a major limitation of the technique, specifically because of the symmetrical nature of the VLC channel \cite{GSSK2}.

In this work, we propose a novel optical modulation scheme that we call amplitude, phase, and quadrant (APQ) modulation. The proposed APQ technique  can be used to transmit high-order modulation signals via a single transmitting LED and a single PD. This is done by  converting the high-order complex constellation symbols into three different components  that carry the amplitude, phase and quadrant information of the symbol. Each of the three components is represented by a unipolar pulse-amplitude modulation (PAM) signal of a suitable order. The three signals are then superimposed in the power domain and sent simultaneously. The receiving terminal performs successive interference cancellation (SIC) to decode and separate the three signals and uses the  amplitude, phase and quadrant components to constitute the complex symbol.

The remainder of the paper is organized as follows: Section \ref{sec:model} describes the channel and system model of an indoor VLC downlink network. Section \ref{sec:benchmark} presents the benchmark model used for comparison and evaluation. Numerical results and related discussions are presented  in Section \ref{sec:results}, while closing remarks are provided in Section \ref{sec:conc}.

\section{Channel and System Model}
\label{sec:model}

\begin{figure}[h]
\center
\includegraphics[width=3.5in,height=2.5in]{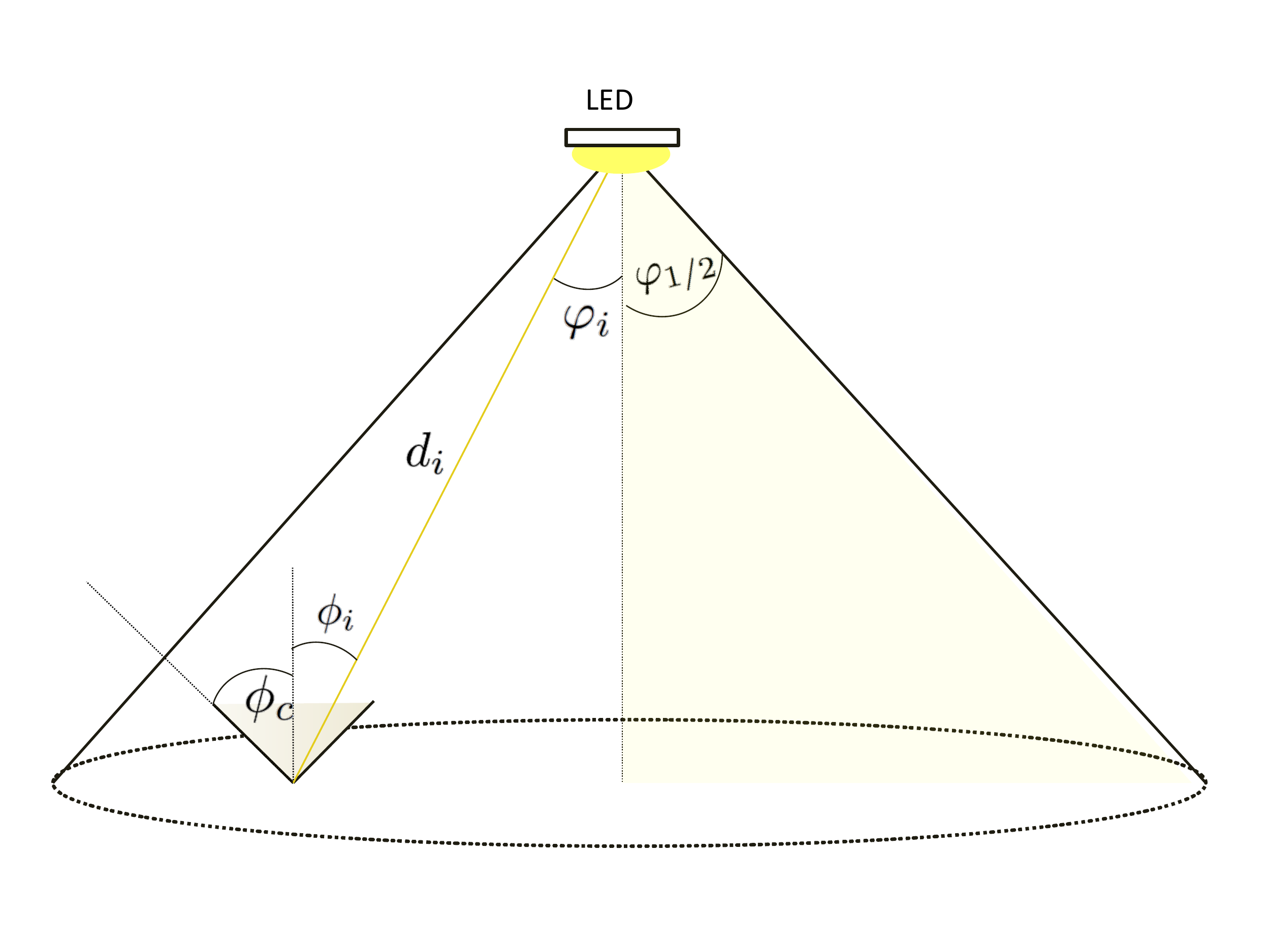}
\caption{VLC channel model.}
\label{fig:system}
\end{figure}

We consider an indoor  downlink VLC system realized by a LOS communication link as shown in Fig. \ref{fig:system}.  The channel gain between the transmitting LED and the receiving PD is given by
\begin{equation}\label{equ:hi}
h= \begin{cases}\frac{A}{d^2}R_o(\varphi)T_s(\phi)g(\phi)\cos(\phi), \qquad & 0\leq\phi\leq \phi_{c}\\ 0,&\phi>\phi_{c},
\end{cases}
\end{equation}
\noindent where  $A$ represents the receiver PD area, $d$ denotes the distance between the LED  and the PD, $\varphi$ is the angle of emergence with respect to the transmitter axis, $\phi$  is the angle of incidence with respect to the receiver axis, $\phi_{c}$ is the field of view (FOV) of the PD, $T_s(\phi)$ is the gain of the optical filter and $g(\phi)$ is the gain of the optical concentrator, which is expressed as

\begin{equation}\label{equ:g}
g(\phi) = \begin{cases} \frac{n^2}{\sin^2(\phi_{c})},\qquad \qquad & 0\leq\phi\leq \phi_{c}\\
 0, \qquad &\phi>\phi_{c},
 \end{cases}
\end{equation}
\noindent where $n$ denotes the corresponding  refractive index. Moreover, $R_o(\varphi)$ in (\ref{equ:hi}) is the Lambertian radiant intensity of the transmitting LEDs, which  can be expressed as
\begin{equation}\label{equ:R_o}
R_o(\varphi) = \frac{(m+1)}{2\pi}\cos^m(\varphi),
\end{equation}
where $m$ is the order of Lambertian emission, calculated as
\begin{equation}\label{equ:m}
 m =\frac{- \ln(2)}{\ln(\cos(\varphi_{1/2}))}
\end{equation}
with $\varphi_{1/2}$ denoting  the transmitter semi-angle at half power.
Moreover, the  noise at the receiving terminal is drawn from   a circularly-symmetric Gaussian distribution of zero mean and variance
\begin{equation}\label{equ:totalnoise}
\sigma_{n}^2 = \sigma_{sh}^2 + \sigma_{th}^2,
\end{equation}
where $\sigma_{sh}^2 $ and $ \sigma_{th}^2$ are the variances of the shot noise and  thermal noise, respectively.

\section{Amplitude, Phase and Quadrature (APQ) Modulation }
In order to transmit an M-ary modulation signal, we categorize each of the M symbols by three parameters, namely, amplitude, phase and quadrant. Each one of these parameters is represented by a unipolar PAM signal, then the three signals are superimposed in the power domain and transmitted simultaneously. As a result, the circular constellation is transformed into three different linear constellations in order  to fit into the constraints  of IM/DD. The idea of power domain superposition is based on assigning different power levels for the different signals, so that the receiving terminal can perform SIC and extract and decode each of the three signals separately. Thus, the proposed APQ scheme can be utilized to transmit high order modulations using  a single LED and a single PD.

\begin{figure}[!t]
\center
\vspace{-2cm}
\includegraphics[width=6 in,height=4.5in]{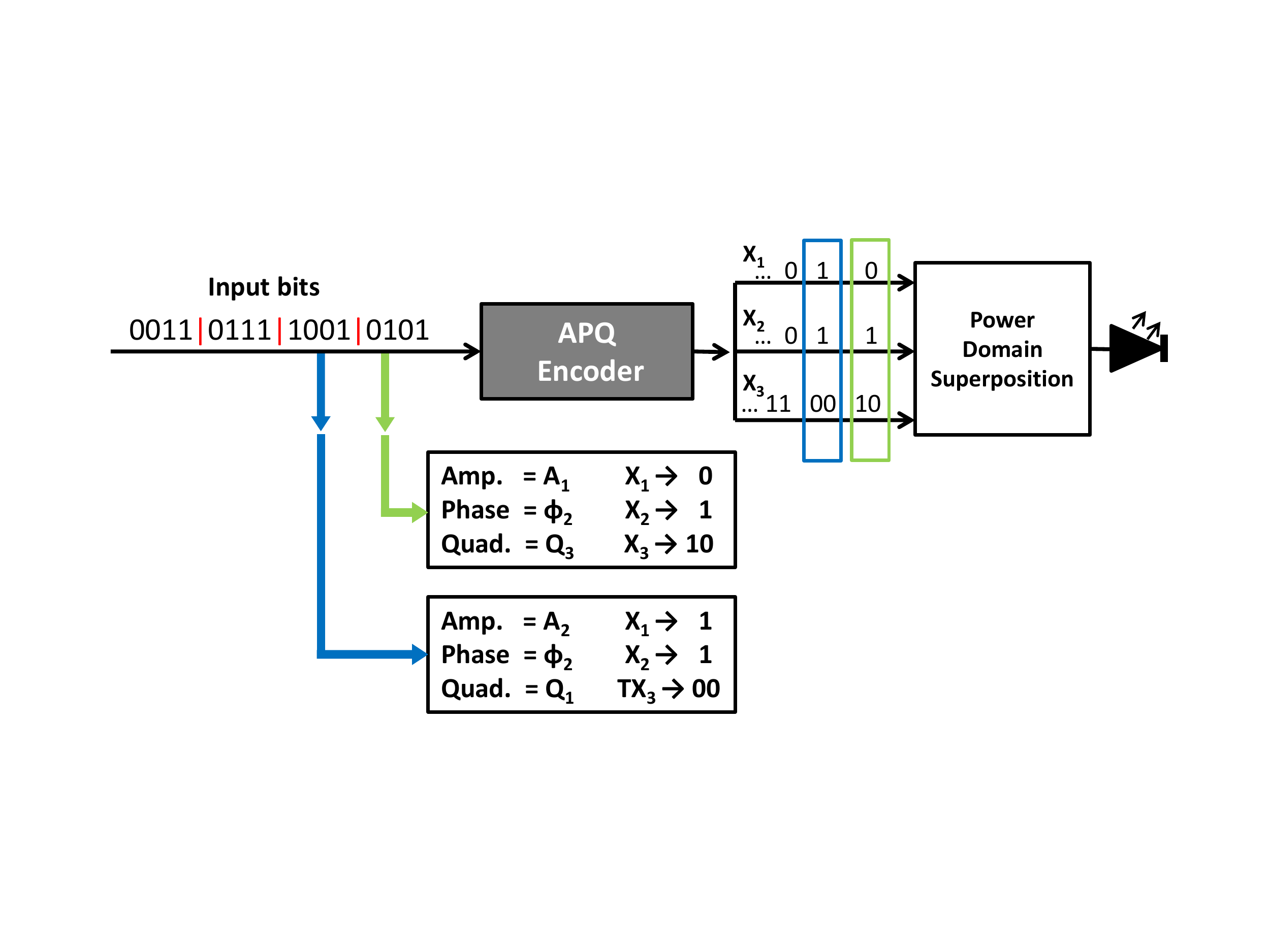}
\vspace{-2cm}
\caption{Illustration of APQ Modulation}
\label{fig:APQ_model}
\end{figure}

\begin{figure}[!t]
\center
\includegraphics[width=3.5in,height=2.5in]{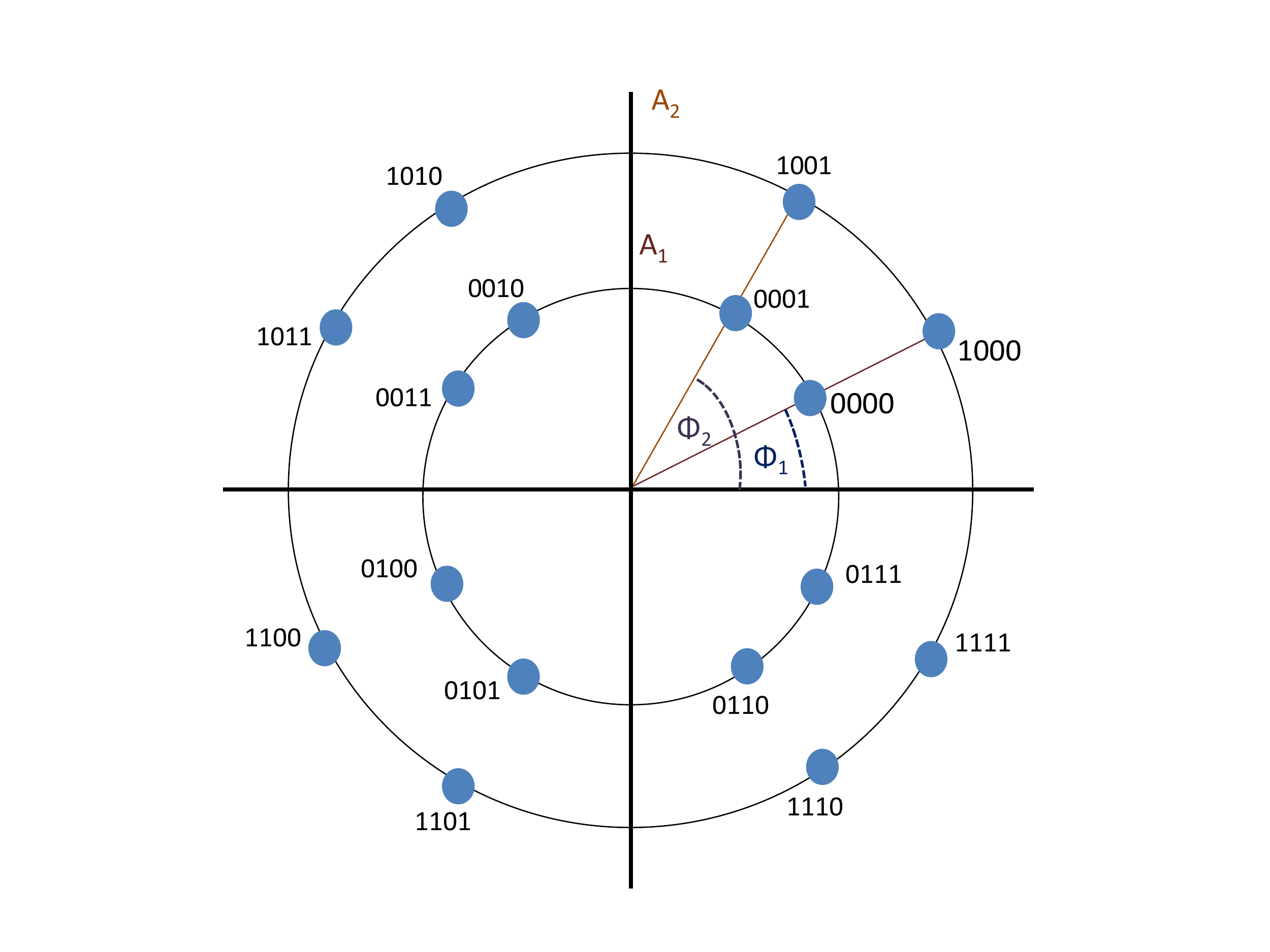}
\caption{Constellation of 16-Ary APSK.}
\label{fig:16constellation}
\end{figure}

For example, the 16 amplitude and phase-shift keying (APSK) constellation shown in Fig. \ref{fig:16constellation}, can be represented by the signals $x_1$, $x_2$ and $x_3$, where   $x_1$  is a 2-PAM signal  representing the two possible  amplitude levels (i.e., $A_1, A_2$), $x_2$ is a 2-PAM signal  representing the two possible phase values ( i.e., $\phi_1, \phi_2$) and  $x_3$ is a 4-PAM signal representing the four different quadrants of the constellation. The APQ encoder is  shown in Fig. \ref{fig:APQ_model}. The three signals are superimposed in the power domain and sent as a single signal. To this end, different power levels are assigned for the three signals to facilitate SIC at the receiver.  We implement simple fixed power allocation (FPA), where the power values are calculated as
\begin{equation}
P_i=\alpha P_{i-1},
\end{equation}
where $0<\alpha<1$ and $\sum_{i=1}^{3} P_i = 1$.
Consequently, the transmitted signal can be written as
\begin{equation}
x=P_1 x_1 + P_2 x_2 + P_3 x_3.
\end{equation}

At the receiving terminal, given that $P_1 > P_2 + P_3$, $x_1$ is directly decoded regarding $x_2$ and $x_3$ as noise. Once decoded, $x_1$ is subtracted from the received signal,  yielding $P_2 x_2 + P_3 x_3$, where $P_2 > P_1$, then $x_2$ can be decoded and subtracted and finally $x_3$ can be decoded.

\section{Error Rate Analysis}
\label{sec:analysis}
The symbol error rate (SER) in APQ modulation can be expressed as:
\begin{equation}
\mbox{Pr}=1-\left( (1-\mbox{Pr}_{1}) \times (1-\mbox{Pr}_{2}) \times (1-\mbox{Pr}_{3}) \right),
\end{equation}
where $\mbox{Pr}_{1}$, $\mbox{Pr}_{2}$ and $\mbox{Pr}_{3}$  are the error probabilities in detecting the amplitude, phase and quadrant signals, respectively.

For the case of 16-ary APQ modulation, $x_1$ and $x_2$ are represented by 2-PAM signals whereas $x_3$ is represented by a 4-PAM signal. The   error probability in detecting the amplitude signal, $x_1$, can be written as
\begin{equation}
\begin{aligned}
\mbox{Pr}_{1} & = \frac{1}{16}\sum_{i=1}^{2^3} \mathcal{Q}\left(\frac{  \gamma h}{\sigma_n} \left(\frac{P_1}{2} -  [P_2 P_3] [\Delta_{2i} \Delta_{3i}]^{T} \right)\right)  \\ & +  \frac{1}{16}\sum_{i=1}^{2^3}  \mathcal{Q}\left(\frac{  \gamma h}{\sigma_n} \left(\frac{P_1}{2} +  [P_2 P_3] [\Delta_{2i} \Delta_{3i}]^{T} \right)\right),
\end{aligned}
\end{equation}
where $\Delta_2=[0 0 0 0 1 1 1 1]^{T}$ and $\Delta_3=[\delta_1 \delta_2 \delta_3 \delta_4 \delta_1 \delta_2 \delta_3 \delta_4]^{T}$ represent all the possible combinations of  the interference caused by $x_2$ and $x_3$, respectively and $\delta_k$ denotes the power level of the $k^{th}$ level in the 4-PAM signal.
After decoding and subtracting $x_1$, the error probability in detecting the phase signal, $x_2$, can be written as
\begin{equation}
\begin{aligned}
\mbox{Pr}_{2} & = \sum_{e_1=-1,0,1} \mathcal{P}(e_1) \mbox{Pr}_{2|e1},
\end{aligned}
\end{equation}
where
\begin{equation}
\begin{aligned}
\mbox{Pr}_{2|e1} & =  \frac{1}{8}  \sum_{i=1}^{2^2} \mathcal{Q}\left(\frac{  \gamma h}{\sigma_n} \left(\frac{P_2}{2} -  P_3 \Delta_{3}-  e_1 P_1 \right)\right) \\ & +  \frac{1}{8} \sum_{i=1}^{2^2}  \mathcal{Q}\left(\frac{  \gamma h}{\sigma_n} \left(\frac{P_2}{2} +  P_3 \Delta_{3} +e_1 P_1\right)\right),
\end{aligned}
\end{equation}
and
\begin{equation}
\begin{aligned}
\mathcal{P}(e_1) = \begin{cases} 1-\mbox{Pr}_{1} & {e_1} = 0 \\ \frac{1}{2}\mbox{Pr}_{1} & {e_i} = -1,1.
 \end{cases}
 \end{aligned}
\end{equation}
Its is noted here that $e_1$ represents the residual interference caused by the detection errors in the first detection stage.

Finally, the error probability in detecting the quadrant signal, $x_3$, can be written as
\begin{equation}
\begin{aligned}\label{equ:BER1}
\mbox{Pr}_{3}& =   \sum_{e_{2= -1,0,1}} \sum_{e_{1= -1,0,1}}    \mathcal{P}(e_2|e_1)  \mathcal{P}(e_1)  \mbox{Pr}_{3|e_1,e_2},
\end{aligned}
\end{equation}
where
\begin{equation}
\begin{aligned}\label{equ:BER1}
\mbox{Pr}_{3|e_1,e_2} & =  \frac{3}{4}  \mathcal{Q}\left(\frac{  \gamma h}{\sigma_n} \left(\frac{P_3}{6} -   e_1 P_1  -   e_2 P_2\right)\right) \\ & +
 \frac{3}{4}  \mathcal{Q}\left(\frac{  \gamma h}{\sigma_n} \left(\frac{P_3}{6} +   e_1 P_1  +   e_2 P_2\right)\right),
\end{aligned}
\end{equation}
and
\begin{equation}
\mathcal{P}(e_2|e_1) = \begin{cases} 1-\mbox{ Pr}_{2|e_1} & {e_2} = 0 \\ \frac{1}{2}\mbox{Pr}_{2|e_1} & {e_2} = -1,1.
 \end{cases}
\end{equation}

\section{Benchmark Model}
\label{sec:benchmark}
For the purpose of comparison and evaluation, we implement the optical GSSK scheme. In GSSK, an M-ary modulation scheme is transmitted using a total of $\log_{2}(M)$ transmitting LEDs. The
number and position of the ones and zeros in the symbol to be transmitted determine which groups of LEDs  transmits during the symbol duration. To this effect, $M$ different combinations of active and idle LEDs are used to represent the different $M$ constellation symbols.  Each LED transmits an on-off keying (OOK) signal, and the channel state information (CSI) available at the receiving terminal is exploited to determine the combination of active LEDs, and thus,  the intended symbol. A detailed analysis and performance evaluation of GSSK can be found in\cite{GSSK1,GSSK2,GSSK3}. It is worth noting that, GSSK is highly dependent on the channel gain differences of the spatially separated LEDs. Consequently, the existence of similar channel gains, which is  typical in VLC due to the channel symmetry,  results in detrimental error performance. In this case,  multiple PDs can be used  to create   channel gain variations at the receiving terminal in order to realize  acceptable  error performance, as demonstrated in \cite{GSSK1}.

\section{Results and Discussions}
\label{sec:results}
In this section, we evaluate the performance of the proposed APQ scheme by considering different setup scenarios, which are based on the system model presented in Section \ref{sec:model}. The obtained results are compared to the benchmark model presented in Section \ref{sec:benchmark} under the same average power constraints to ensure comparability.  We consider a $4 \times 4 \times 3$ m$^3$ room with a maximum of five high brightness white LEDs and a single PD. We assume that the transmitting LEDs are located at  a height of $ z = 2.50$ m, while the receiving PDs are placed at a height of $z = 0.75$ m.
The used simulation parameters are shown in Table \ref{Tab:Parameters}, while the front-ends' coordinates are shown in Table \ref{Tab:Txs}.

\begin{figure}[!t]
\center
\includegraphics[width=4in,height=3.5in]{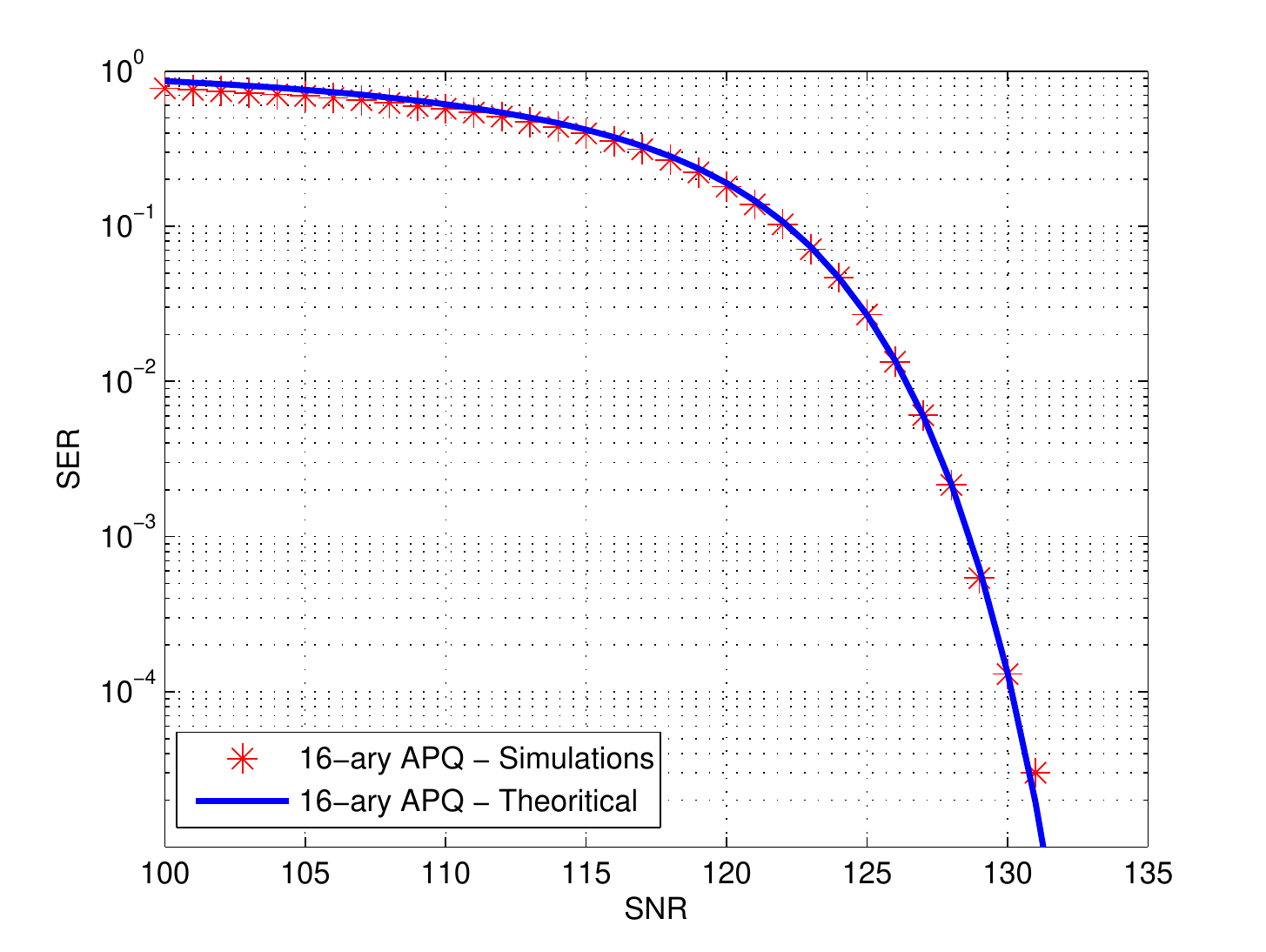}
\caption{SER Performance vs Transmit SNR.}
\label{fig:perfectCSI}
\end{figure}

First, we validate the SER expression derived in Section \ref{sec:analysis}. Fig. \ref{fig:perfectCSI} illustrates the SER performance of 16-ary APQ scheme with regard to the transmit signal-to-noise-ratio (SNR). Since the channel gain is in the order of $10^{-4}$, the corresponding results   exhibit an offset of about  $80$ dB  with respect to the  SNR at the receiver site. It is shown that the derived analytical results are in excellent agreement with the respective Monte Carlo simulation results.
\begin{table}[!t]
\small
\caption{Simulation Parameters}
\center
\begin{tabular}{|l| l| l|}
\hline
\textbf{Description}&\textbf{Notation}&\textbf{Value} \\ \hline
LED power&$P_{\rm LED}$& 0.25 W\\
Transmitter semi-angle&$\varphi_{i}$& $30^{\circ}$ \\
FOV of the PDs&$\phi_{c_i}$& $30^{\circ}$ \\
Physical area of PD&$A_i$ &$1.0$ ${\rm cm^2}$ \\
Refractive index of PD lens&$n$&$ 1.5$ \\
Gain of optical filter&$T_s (\phi_{li})$&$ 1.0$ \\
Data rate&$B$&$10$ ${\rm Mbps}$\\
\hline
\end{tabular}
\label{Tab:Parameters}
\end{table}

\begin{table}[!t]
\small
\caption{Locations of Transmitting LEDs}
\center
\begin{tabular}{|l| l ||l| l |}
\hline
\textbf{LED}&\textbf{Coordinate} &\textbf{PD}&\textbf{Coordinate} \\ \hline
Tx1 & (2-d,2-d,3)& Rx1 & (1,1,0.75) \\
Tx2 & (2+d,2+d,3)& Rx2 & (2.1,2.2,0.75)\\
Tx3 & (2,2,3)& Rx3 & (2,2,0.75)\\
Tx4 & (2-d,2+d,3)& &\\
Tx5 & (2+d,2-d,3)& &\\
\hline
\end{tabular}
\label{Tab:Txs}
\end{table}
\vspace{-0.1cm}

\begin{figure}[!t]
\centering
\normalsize
\subfloat[]\centering{\label{ref_label1}\includegraphics[width=2.3in,height=2.2in]{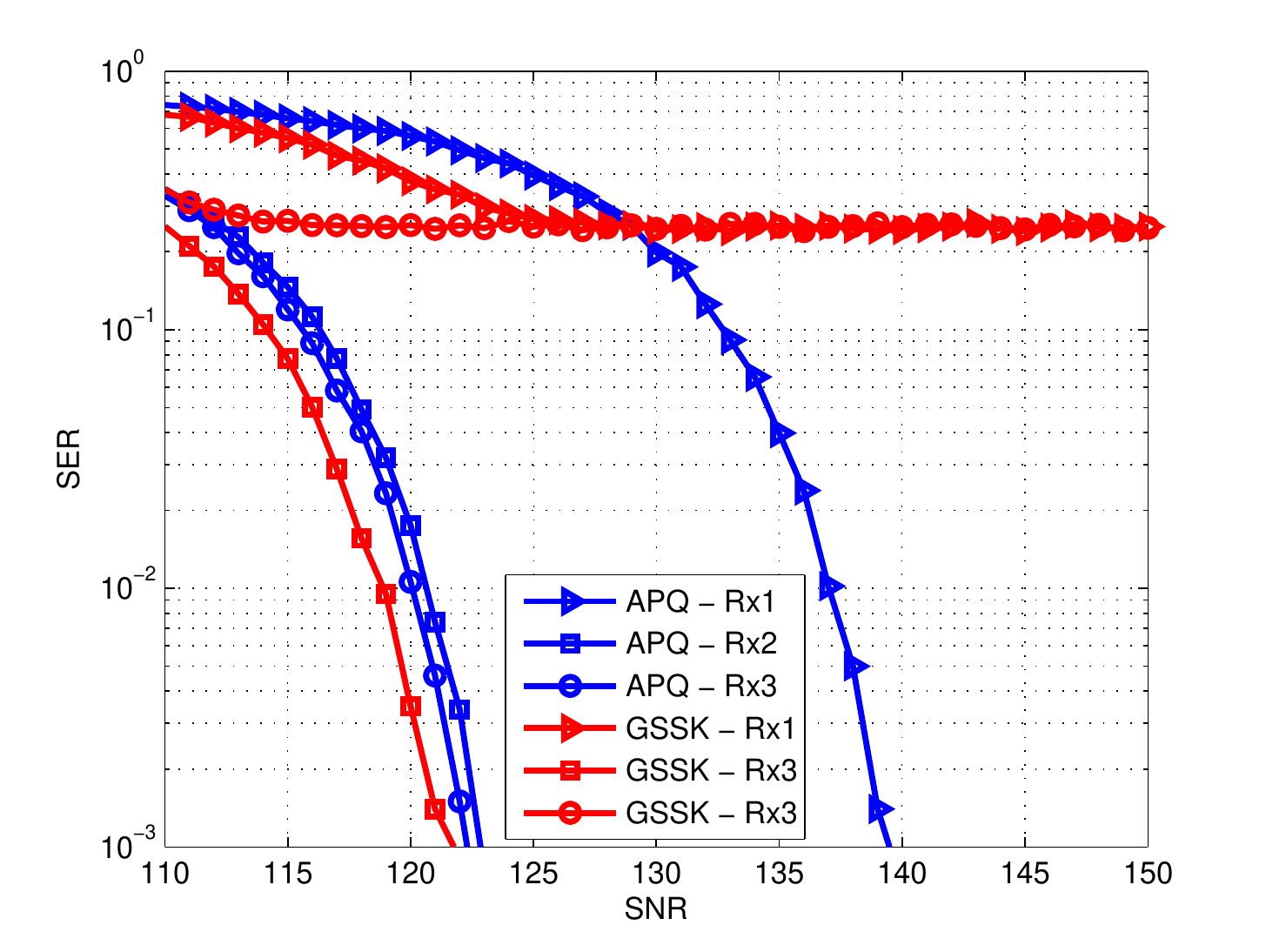}}
\subfloat[]\centering{\label{ref_label2}\includegraphics[width=2.3in,height=2.2in]{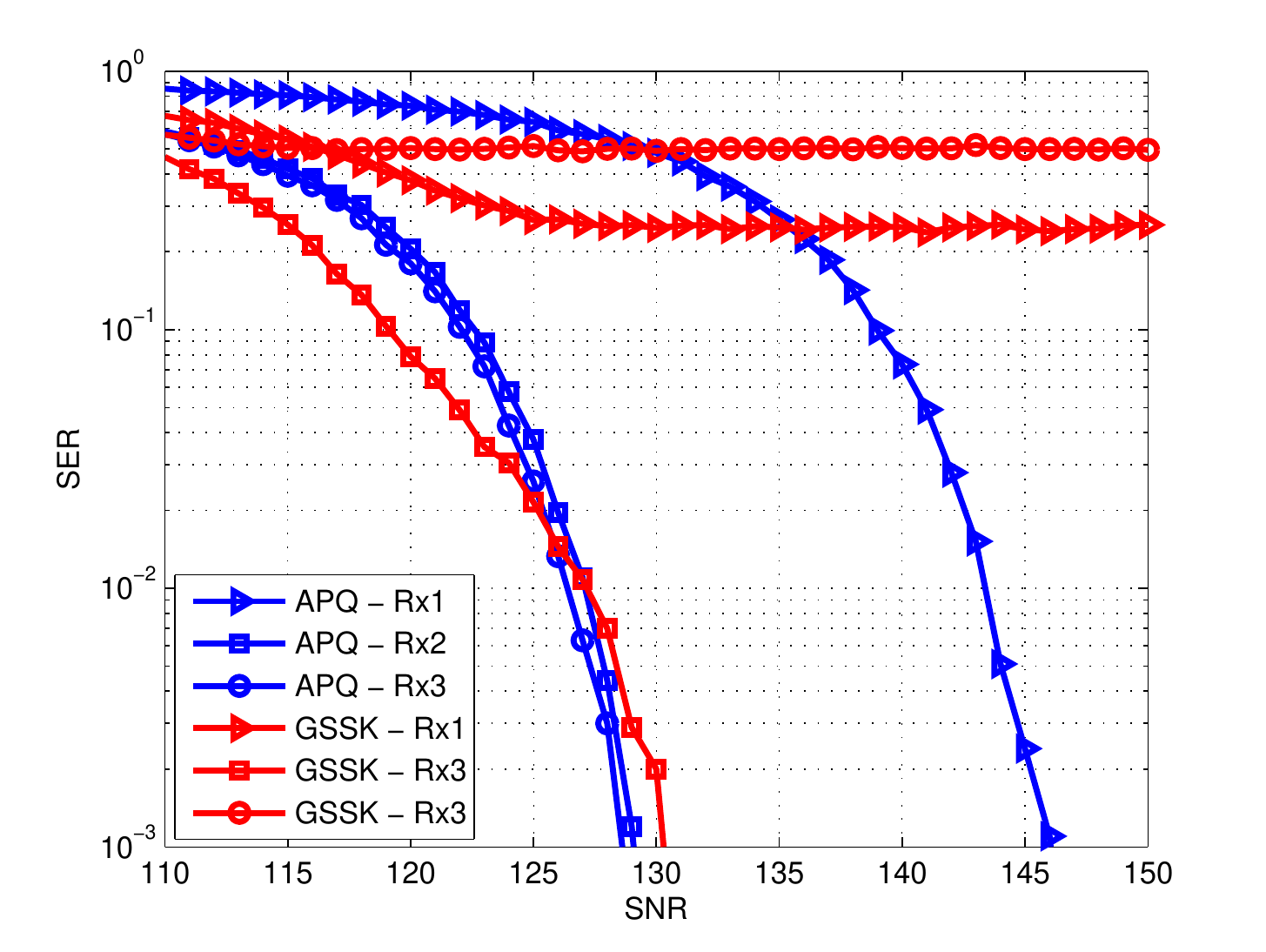}}
\subfloat[]\centering{\label{ref_label2}\includegraphics[width=2.3in,height=2.2in]{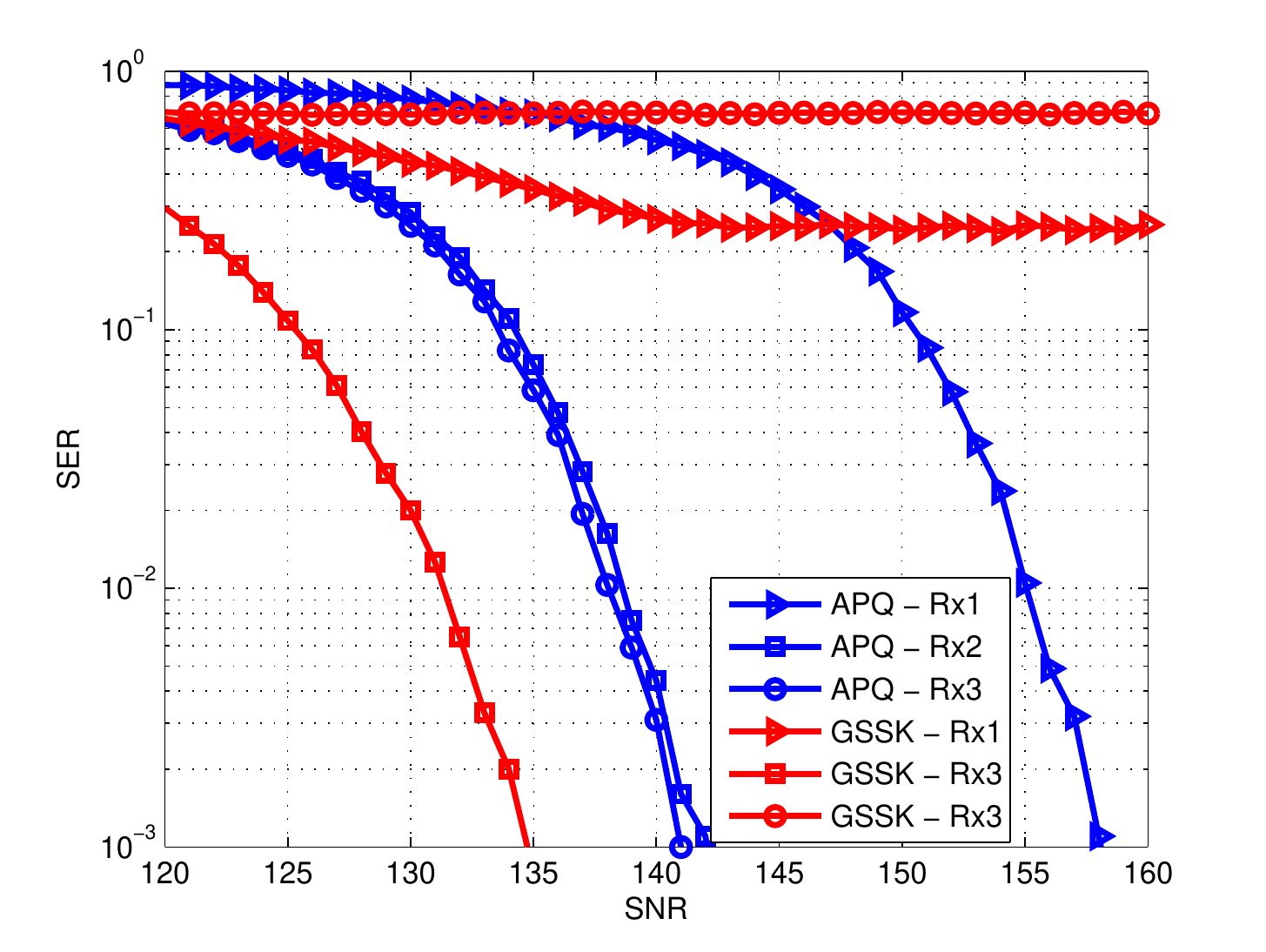}}
\caption{\label{fig:theta30}SER vs Transmit SNR for $d=1$ m   (a) 3 bits/symbol, (b) 4 bits/symbol and (c) 5 bits/symbol.}
\label{Fig:scenario2}
\end{figure}

Next, we examine the SER performance of the proposed APQ scheme, and compare it to the performance of GSSK when the distance between the GSSK transmitters on the x- and y-axis is $d=1$ m. This setup, adopted in \cite{GSSK1}, presents a best case scenario for GSSK, as the transmitters are widely spaced resulting in considerable channel gain differences. On the basis of this setup, we simulate the SER performance for  various static  locations of the receiving PD, i.e., when the PD is located close to the room boundaries, close to the room center, and exactly in the center of the room. The different locations are denoted by Rx1, Rx2 and Rx3 respectively. Fig. \ref{Fig:scenario2} shows that, while GSSK outperforms APQ for the scenario when the receiving PD is located in the proximity of the room center, the proposed APQ scheme provides a good performance regardless of the location of the PD.  It is also evident that GSSK fails to provide meaningful communication link when the receiving PD is located in the room center or at the boundaries.  This is due to the identical channel gains caused by symmetry, that diminishes the  spatial modulation.
\par
\begin{figure}[!t]
\centering
\normalsize
\subfloat[]\centering{\label{ref_label1}\includegraphics[width=2.3in,height=2.2in]{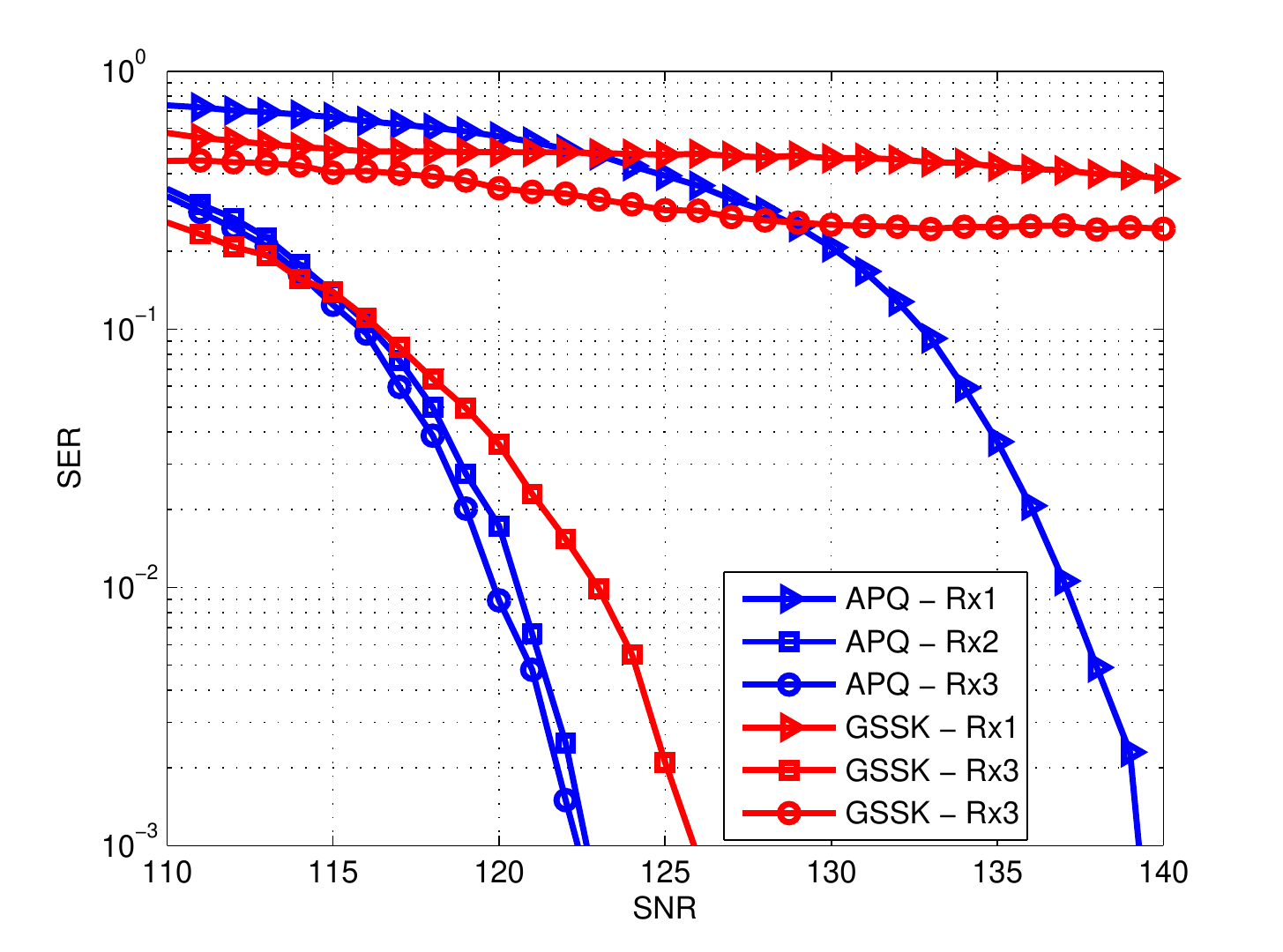}}
\subfloat[]\centering{\label{ref_label2}\includegraphics[width=2.3in,height=2.2in]{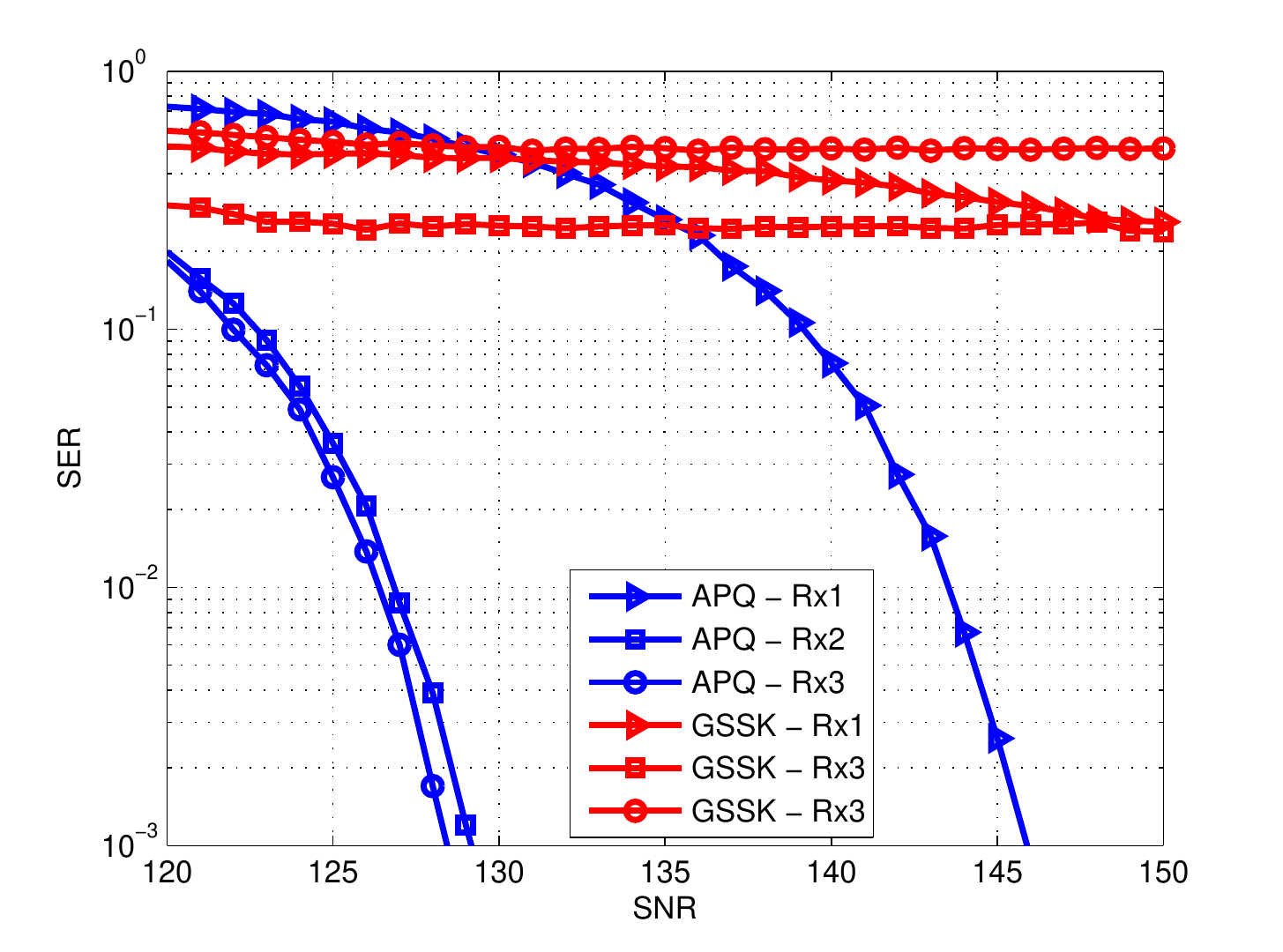}}
\subfloat[]\centering{\label{ref_label2}\includegraphics[width=2.3in,height=2.2in]{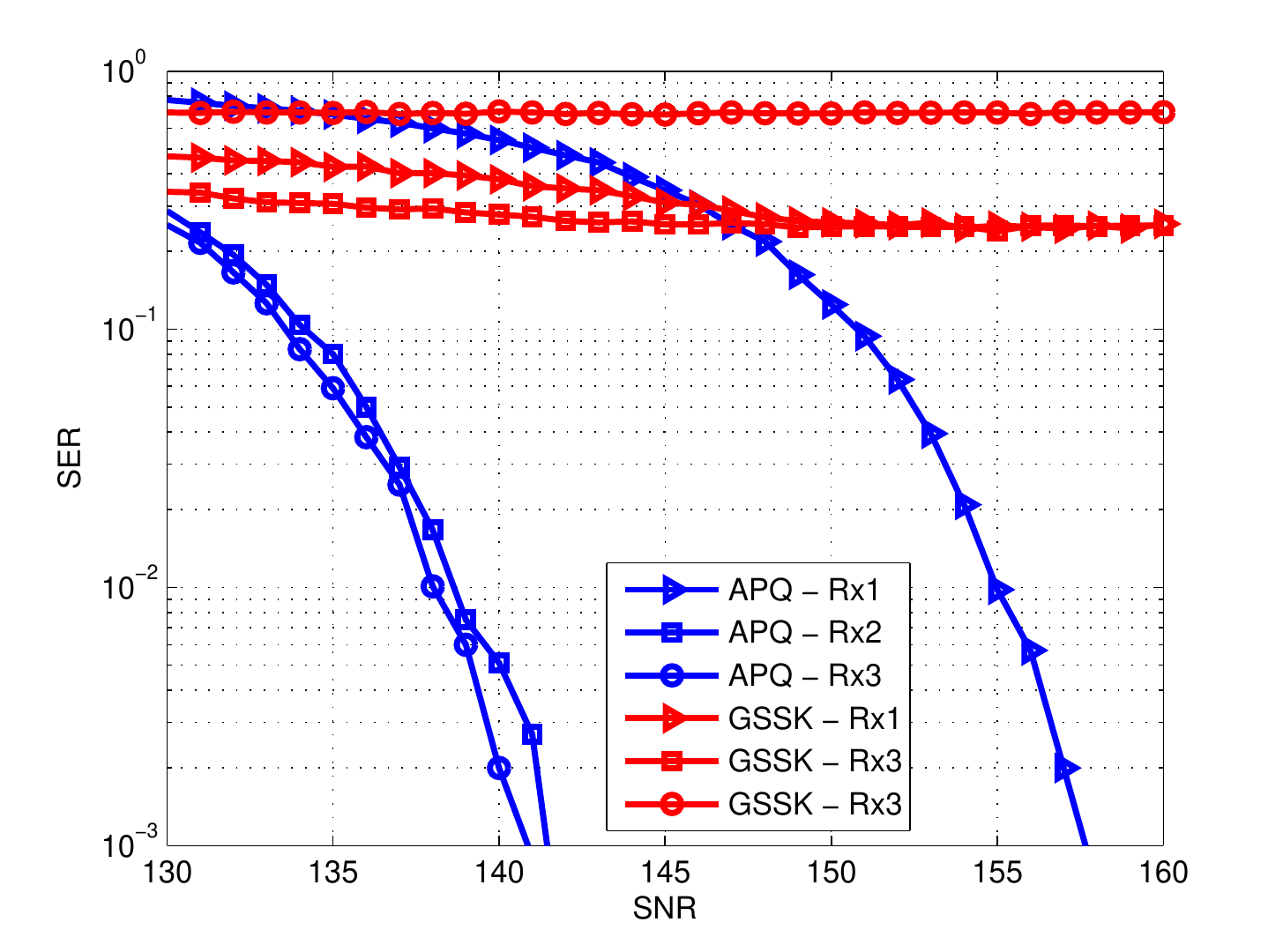}}
\caption{\label{fig:theta30}SER vs Transmit SNR for $d=0.1$ m   (a) 3 bits/symbol, (b) 4 bits/symbol and (c) 5 bits/symbol.}
\label{Fig:scenario1}
\end{figure}

 Next, we examine the effect of the  LEDs' locations on the performance of GSSK. To this end,  we simulate the SER for $d=0.1$ m, which is a typical spacing  that has been considered in \cite{MIMO_VLC_Haas}. Fig. \ref{Fig:scenario1} demonstrates the effect of similar channel gains on the performance of GSSK, and shows that the proposed APQ scheme provides superior performance compared to GSSK under the existence of similar channel gains.

\begin{figure}[!t]
\centering
\normalsize
\subfloat[]\centering{\includegraphics[width=5cm]{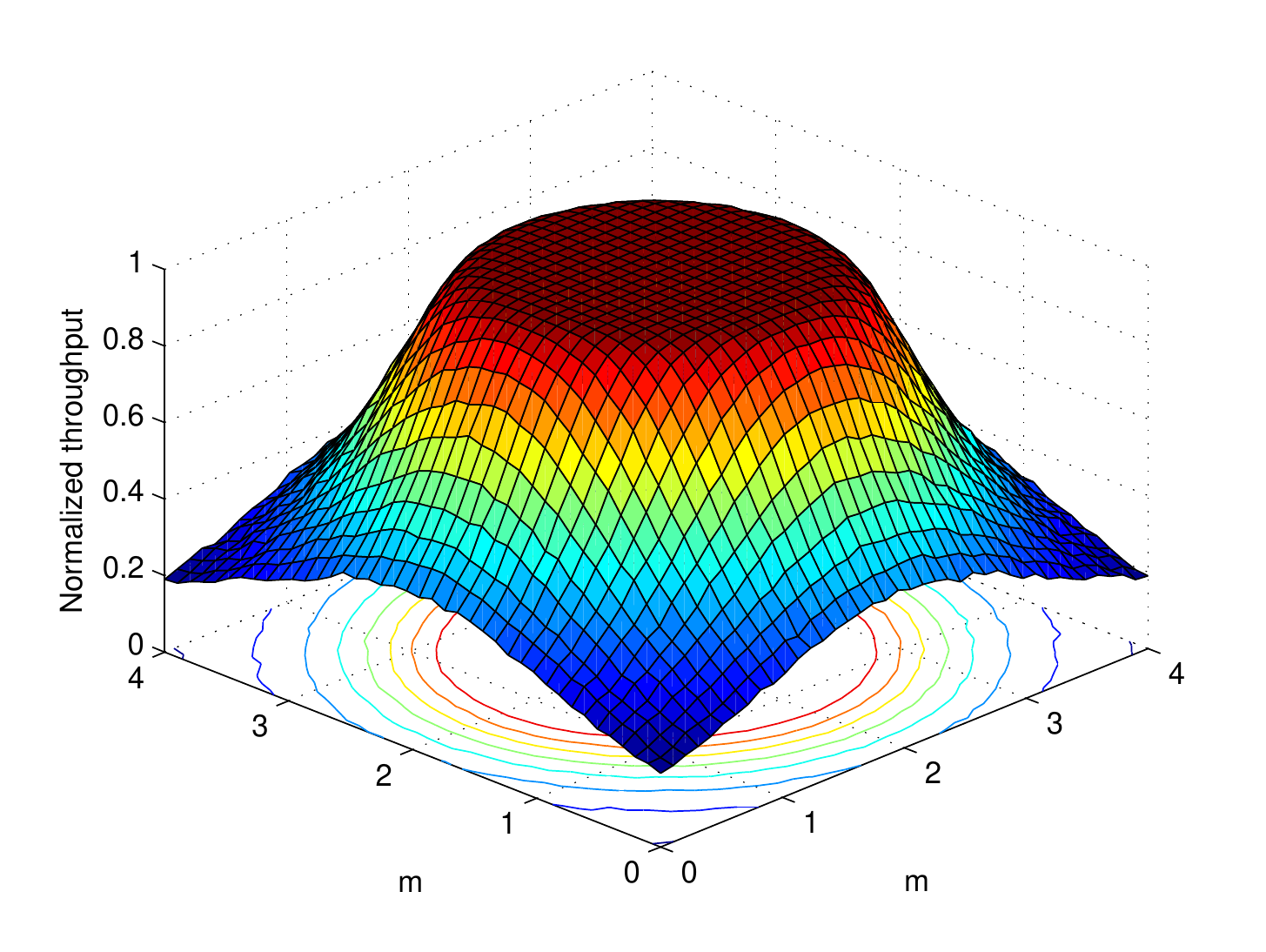} } \subfloat[]\centering{\includegraphics[width=5cm]{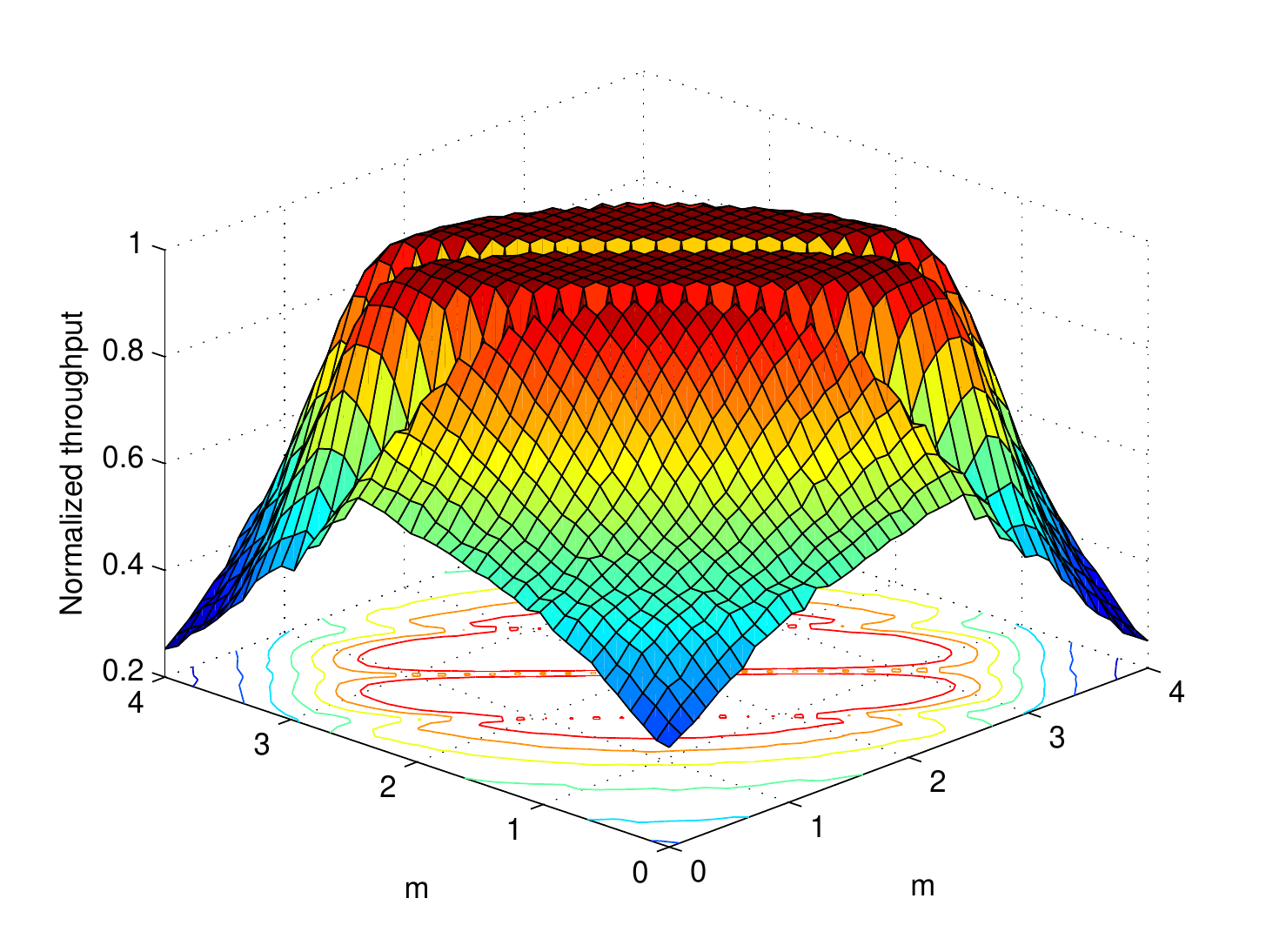}}\\
\subfloat[]\centering{\includegraphics[width=5cm]{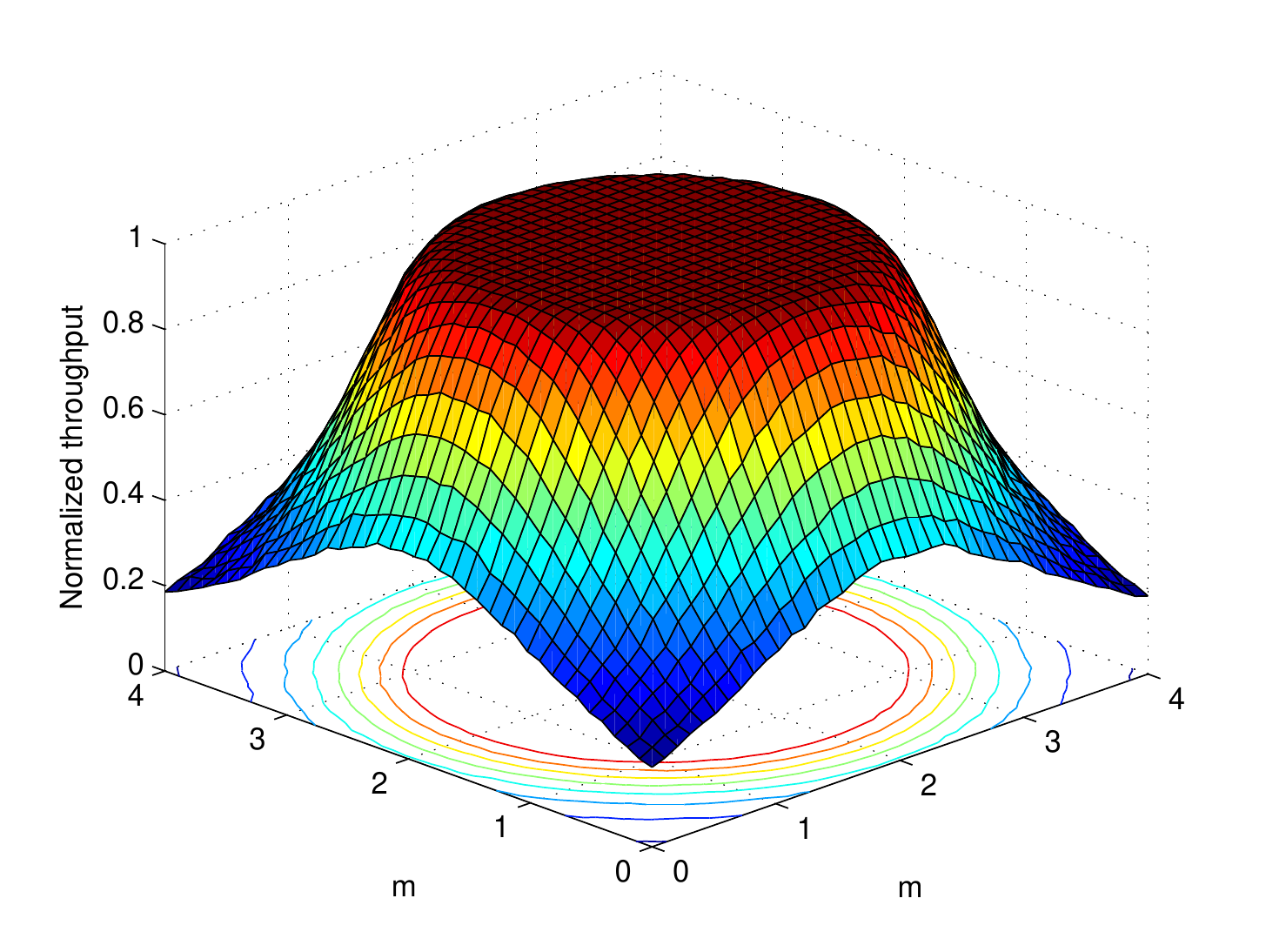}} \subfloat[]\centering{\includegraphics[width=5cm]{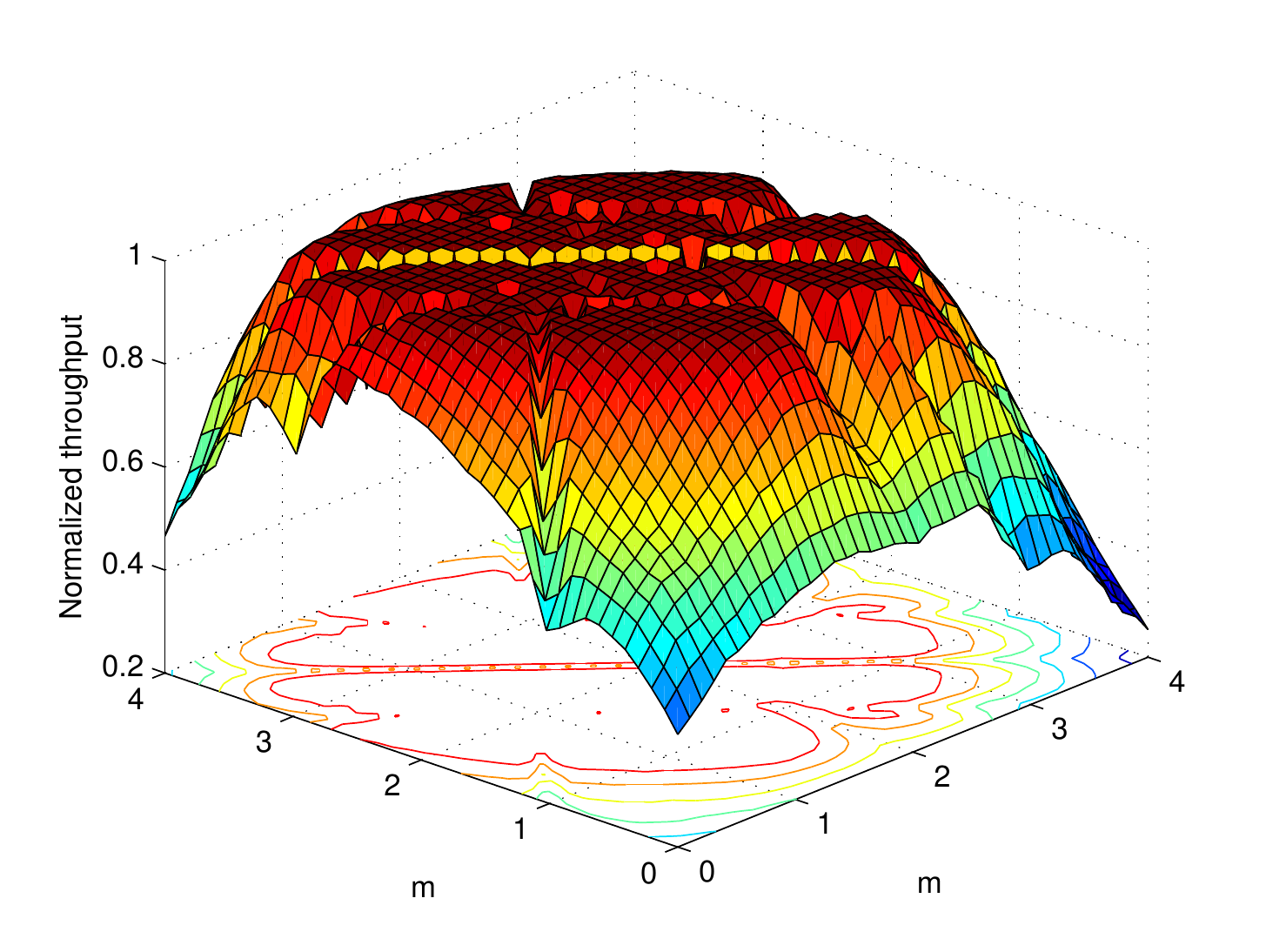}}\\
\subfloat[]\centering{\includegraphics[width=5cm]{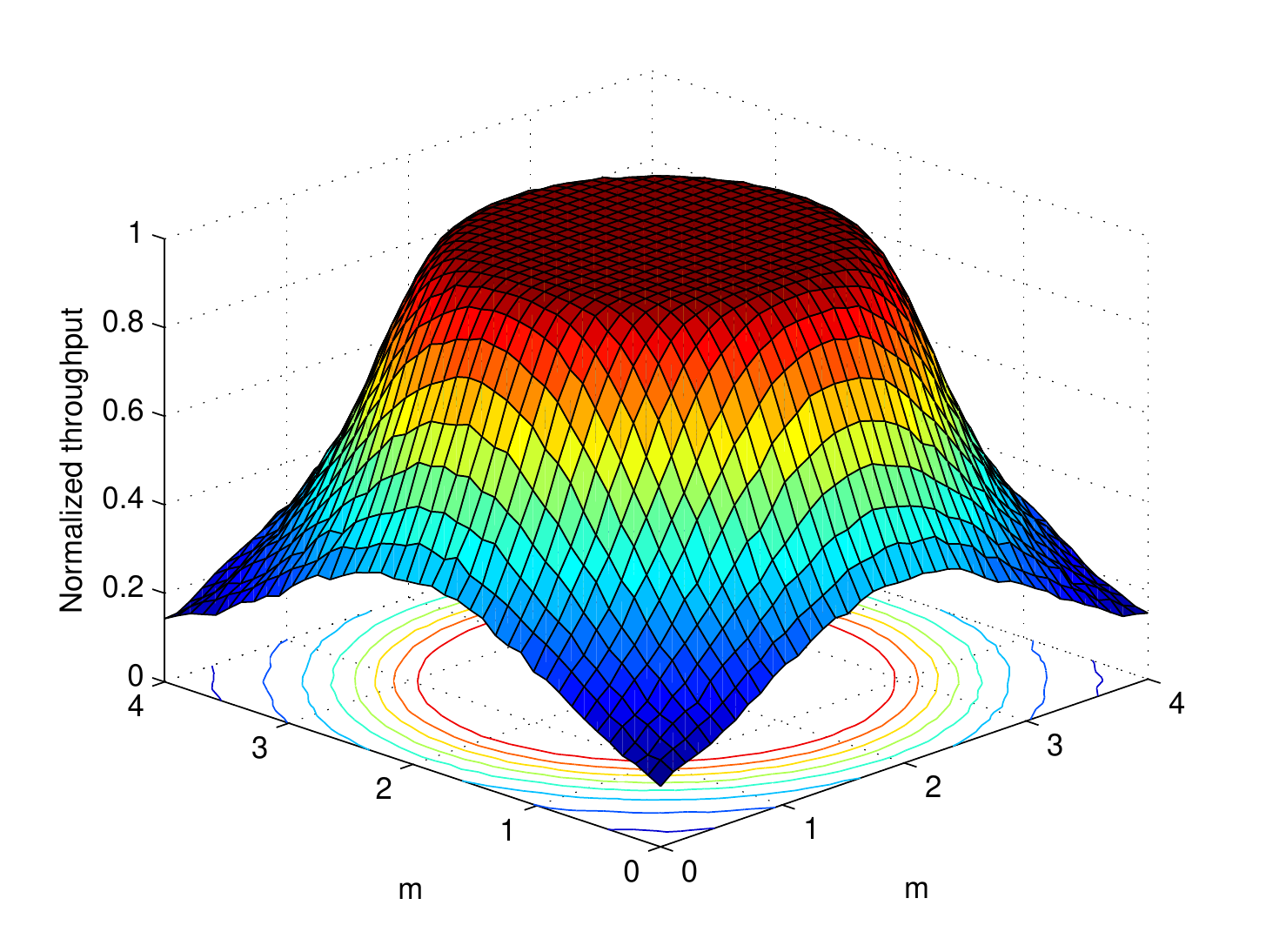}}\subfloat[]\centering{\includegraphics[width=5cm]{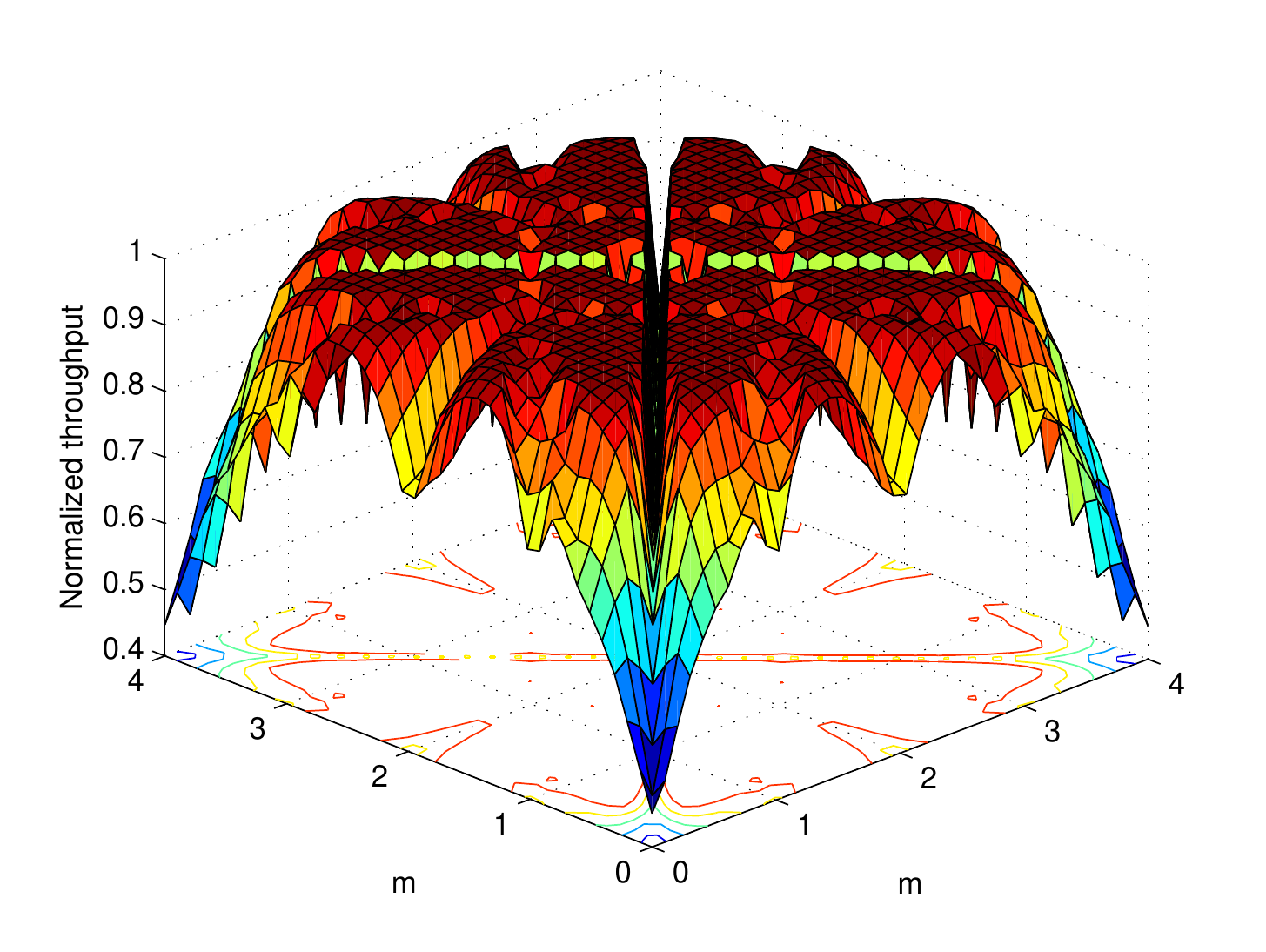}}
\caption{\label{fig:theta15}Normalized throughput across the simulation area, $d=1$ m   for (a) APQ 3 bits/symbol,  (b) GSSK 3 bits/symbol, (c) APQ 4 bits/symbol, (d)  GSSK 4 bits/symbol, (e) APQ 5 bits/symbol, and (f) GSSK 5 bits/symbol.}
\label{Fig:ThScenario2}
\end{figure}

\begin{figure}[h]
\centering
\normalsize
\subfloat[]\centering{\includegraphics[width=5cm]{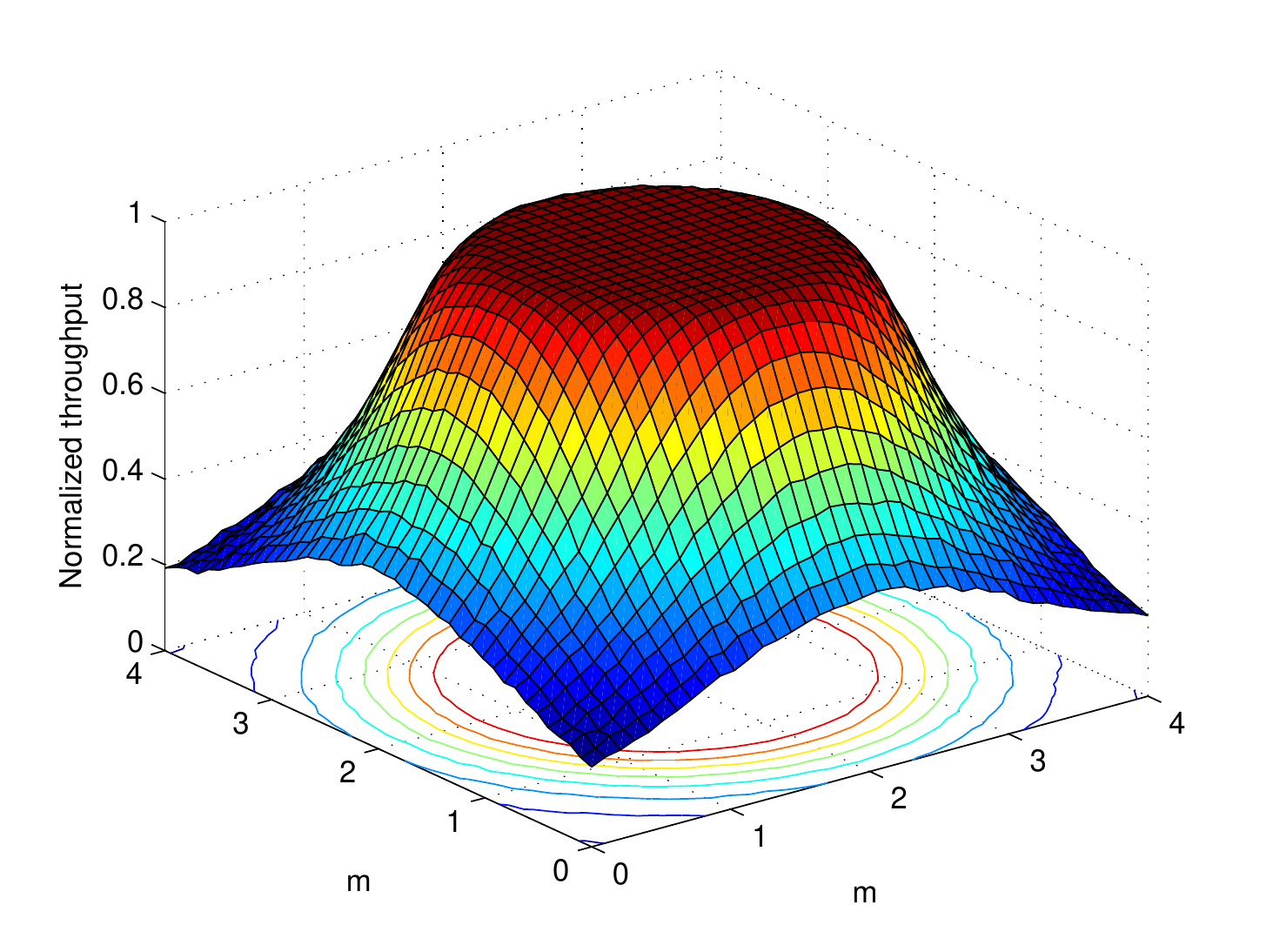} } \subfloat[]\centering{\includegraphics[width=5cm]{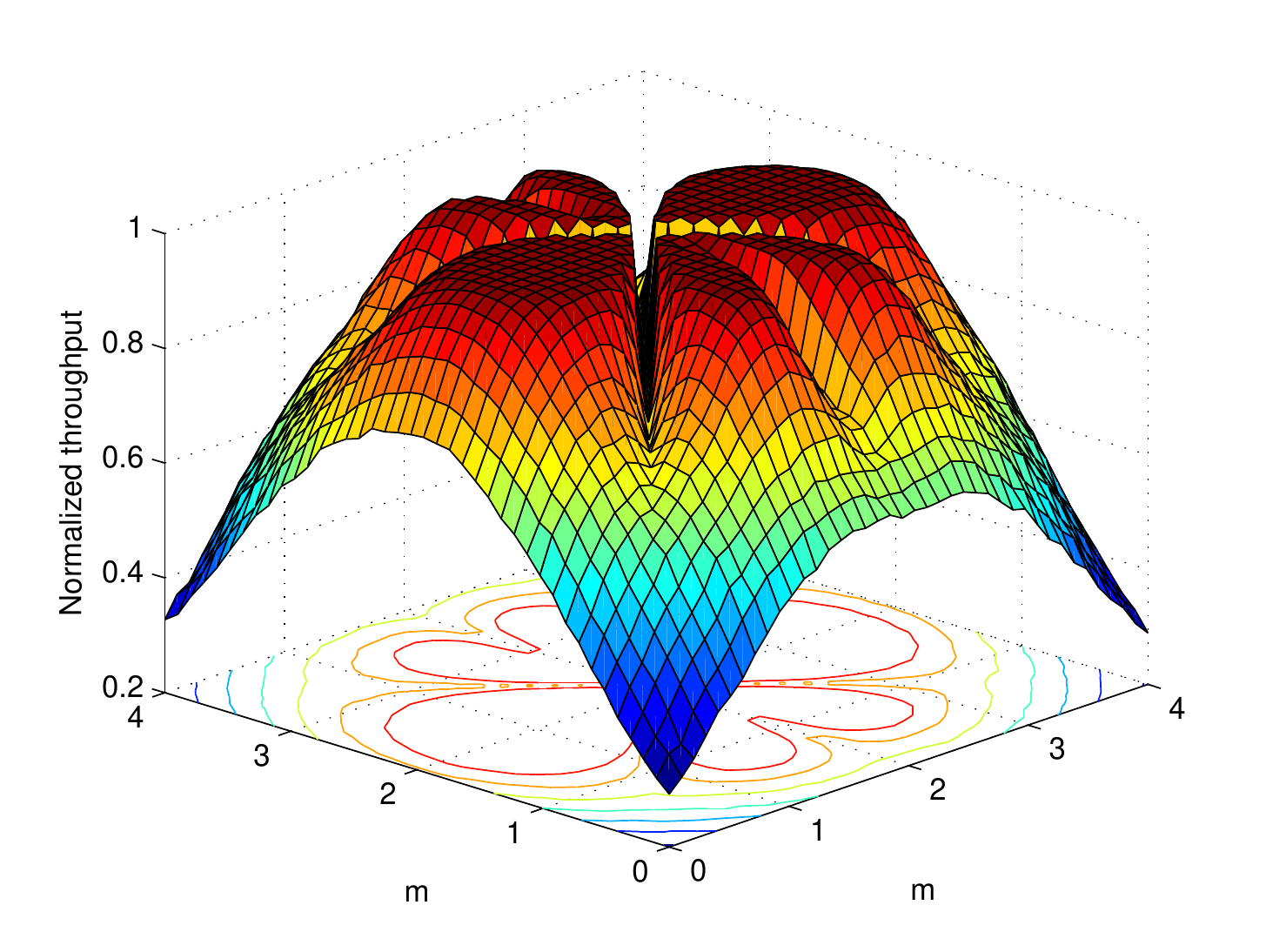}}\\
\subfloat[]\centering{\includegraphics[width=5cm]{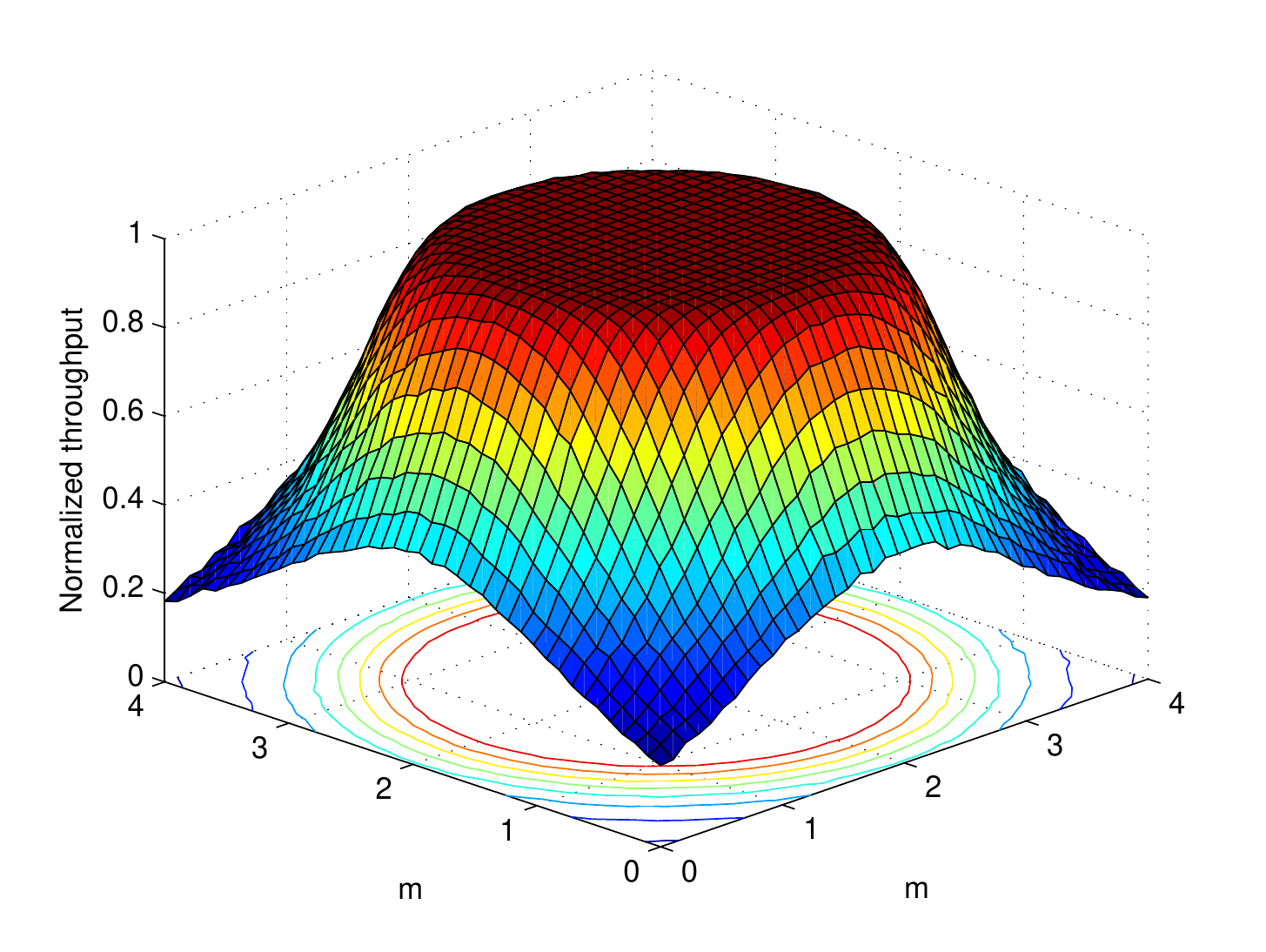}} \subfloat[]\centering{\includegraphics[width=5cm]{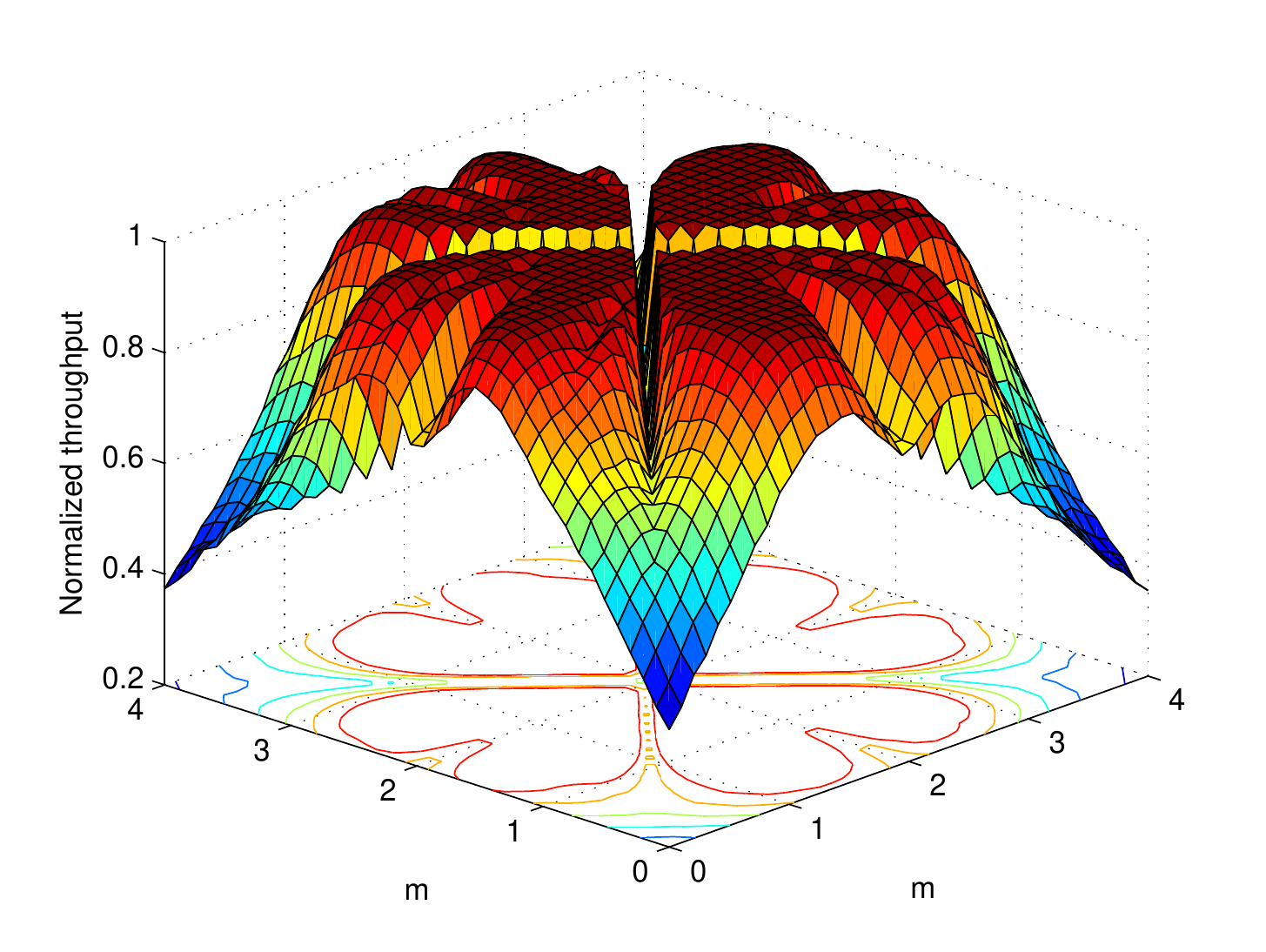}}\\
\subfloat[]\centering{\includegraphics[width=5cm]{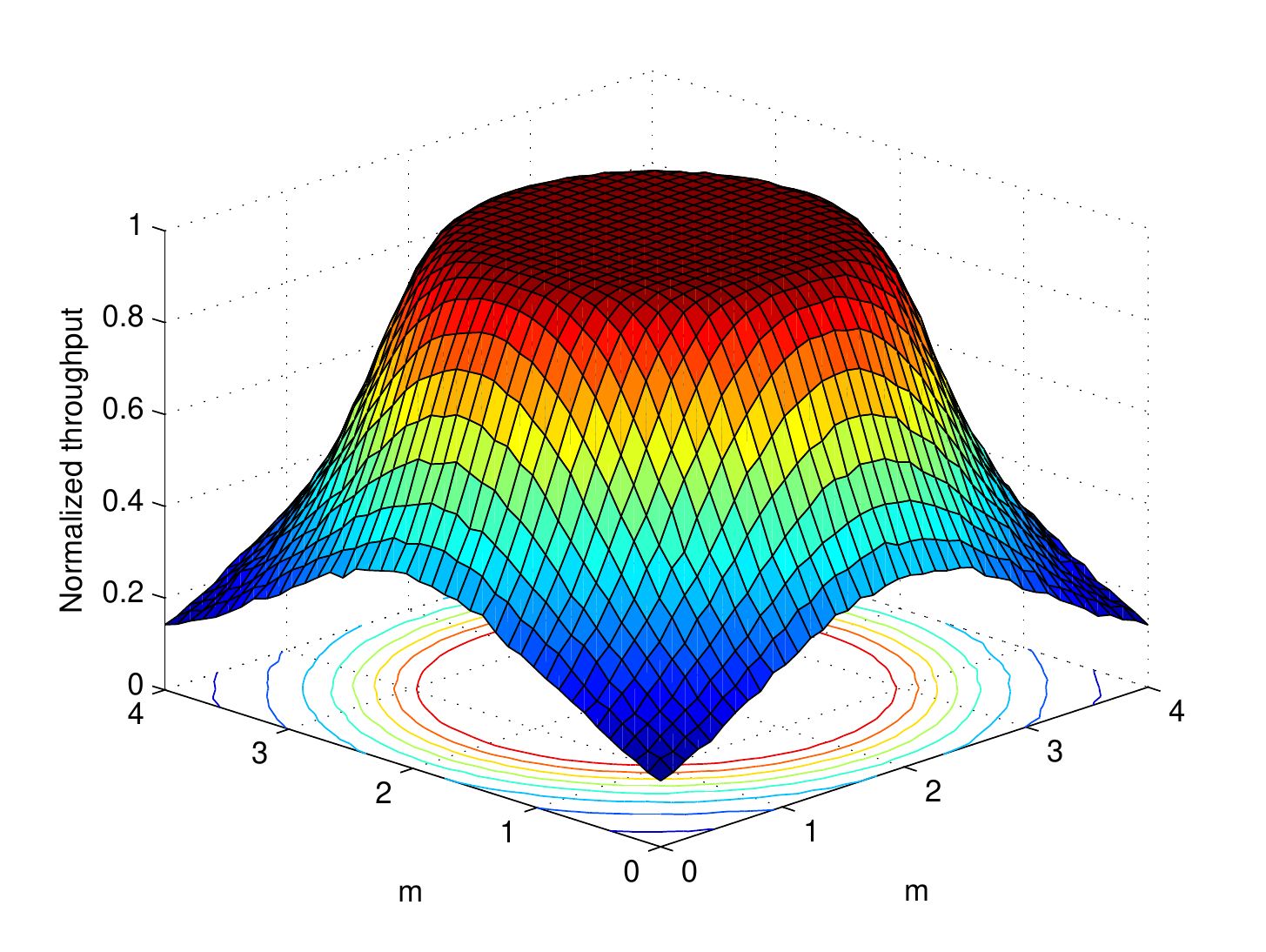}}\subfloat[]\centering{\includegraphics[width=5cm]{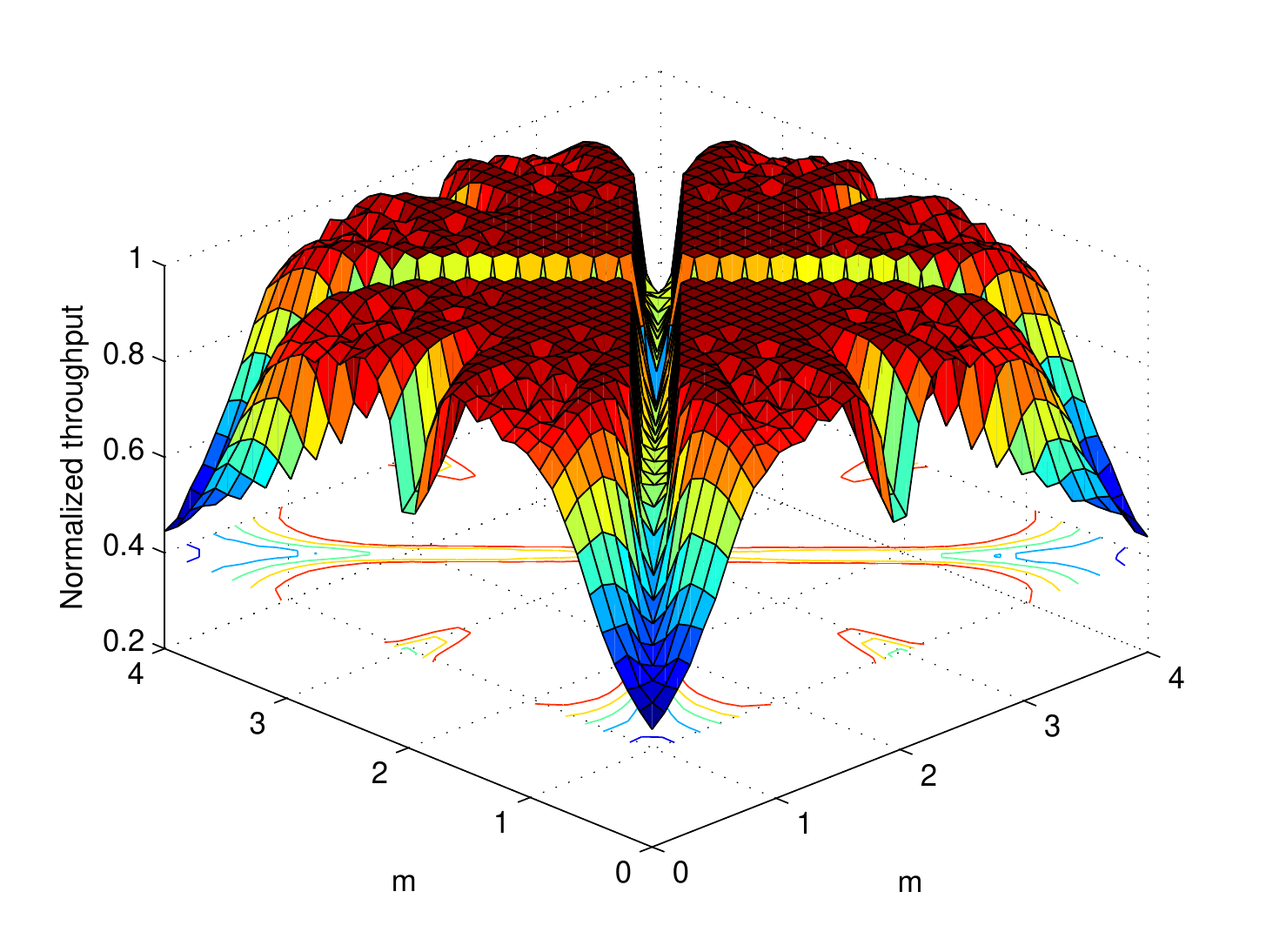}}
\caption{\label{fig:theta15}Normalized throughput across the simulation area,  $d=0.1$ m  for (a) APQ 3 bits/symbol,  (b) GSSK 3 bits/symbol, (c) APQ 4 bits/symbol, (d)  GSSK 4 bits/symbol, (e) APQ 5 bits/symbol, and (f) GSSK 5 bits/symbol.}
\label{Fig:ThScenario1}
\end{figure}

Next, we investigate the achievable system throughput across the simulation area. Fig. \ref{Fig:ThScenario2} and \ref{Fig:ThScenario1} show  the normalized achievable throughput for APQ and GSSK for $d=1$ m and $d=0.1$ m, respectively. The results are obtained for the case of transmitting 3 bits/symbol, 4 bits/symbol and 5 bits/symbol, under fixed transmit SNR of $130$ dB, $140$ dB and $150$ dB respectively.  It is evident that APQ achieves uniform throughput levels across the majority of the simulation area, while a throughput degradation occurs at the boundaries. This is an expected behaviour as the user is receiving from a single LED, which is considered a small cell that serves users within its proximity. On the other hand,  GSSK suffers  huge throughput non-uniformity  due to the similarities between the channel gains observed by the receiving terminal. This throughput non-uniformity  hinders the reliability of the communication link, as the user may encounter a substantial performance loss while moving in the coverage area  of the transmitting LEDs.

\section{Conclusions}
\label{sec:conc}
In this paper, we proposed a novel APQ modulation scheme that is specifically tailored to the requirements imposed  by IM/DD based VLC systems. The proposed scheme transmits high order modulation signals by decomposing the amplitude, phase and quadrant information of each symbol, and transmitting the obtained information simultaneously in the power domain. We demonstrated the transmission of  8-ary, 16-ary and 32-ary signals using a single LED and a single photo detector, thus, the proposed APQ scheme provides a cost-effective solution to improve the achievable data rates in indoor VLC systems.  On the other hand, GSSK requires $N_T$ transmitting LEDs to transmit a $2^{N_T}$-ary signal. Moreover, APQ offers a reliable communication link regardless of the location of the receiving terminal, which makes it suitable for the practical scenarios where the user moves in the proximity of the transmitting LED, while GSSK suffers significant performance degradation when the receiving PD receives similar channel gains from the different transmitting LEDs.

\bibliographystyle{IEEEtran}

\bibliography{APQ_PhyComm}

\end{document}